\documentclass[structabstract]{aa}
\usepackage{txfonts}
\usepackage{graphicx}
\usepackage{natbib}
\usepackage{float}
\begin{document}

\title{Physical properties of the jet from DG Tauri\\ 
on sub-arcsecond scales with HST/STIS
	\thanks{Based on observations made with the 
NASA/ESA {\em Hubble Space Telescope}, obtained at 
the Space Telescope Science Institute, 
which is operated by the Association 
of Universities for Research in Astronomy,
Inc., under NASA contract NAS5-26555.}
}

\author{  L. Maurri\inst{1} 
	\and F. Bacciotti \inst{2}
      \and L. Podio \inst{2}
	 \and J. Eisl\"{o}ffel\inst{3} 
          \and T. P. Ray\inst{4}
           \and R. Mundt\inst{5}
            \and U. Locatelli\inst{6}
	     \and D. Coffey \inst{7}
               }

\institute{Universita' di Firenze, Sez. di Astronomia, Largo E. Fermi 2, 
I-50125 Florence, Italy
   \and I.N.A.F.-Osservatorio Astrofisico di Arcetri, Largo E. Fermi 5,
I-50125  Florence, Italy
\and Th\"uringer Landessternwarte Tautenburg,
Sternwarte 5, D-07778 Tautenburg, Germany
\and School of Cosmic Physics, 
Dublin Institute for Advanced Studies,
31 Fitzwilliam Place, Dublin 2, Ireland
\and Max-Planck-Institut f\"ur Astronomie, K\"onigstuhl 17,
D-69117  Heidelberg, Germany
\and Dipartimento di Matematica,
Universit\`a degli Studi di Roma ``Tor Vergata'',
Via della Ricerca Scientifica 1, 00133 Roma, Italy,
\and School of Physics,
University College Dublin,
 Belfield, Dublin 4, Ireland.
}
  
\date{Received / Accepted}

\abstract 
{Stellar jets are believed to play a key role in star formation,
but the question of how they originate is still being debated. 
}
{
We derive the physical properties at the base of the jet from DG Tau both along and across the flow and as a function of velocity. 
}
{
We analysed seven optical spectra of the DG Tau jet, taken with the Hubble Space 
Telescope Imaging Spectrograph. The spectra were obtained by placing a long-slit parallel to the jet axis and stepping it across the jet width. The resulting position-velocity diagrams in optical forbidden emission lines allowed access to plasma conditions via calculation of emission line ratios. In this way, we produced a 3-D map (2-D in space and 1-D in velocity) of the jet's physical parameters i.e. electron density n$_e$, hydrogen ionisation fraction x$_e$, and total hydrogen density n$_H$. The method used is a new version of the BE-technique. 
}
{
A fundamental improvement is that the new diagnostic method allows us to overcome 
the upper density limit of the standard [\ion{S}{ii}] diagnostics.
As a result, we find at the base of the jet  high electron density, $n_e \sim $ 10$^5$, and  very low ionisation, $x_e \sim 0.02-0.05$, which combine to give a total density up to $n_H \sim $ 3 10$^6$. 
This analysis confirms previous reports of variations in plasma parameters along the jet, (i.e. decrease in density by several orders of magnitude, increase of $x_e$ from 0.05 to a plateau at 0.7 downstream at 2$''$ from the star). Furthermore, a spatial coincidence is revealed between sharp gradients in the 
total density 
and supersonic velocity jumps. 
This strongly suggests that the emission is caused by shock excitation. 
No evidence was found of variations in the parameters across the jet, within a given velocity interval. 
The position-velocity diagrams indicate the presence of both  fast accelerating gas and slower, less collimated material.
We derive the mass outflow rate, $\dot{M}_j$, in the blue-shifted lobe in different velocity channels, that contribute to a total of
$\dot{M}_j \sim$  8 $\pm$ 4 10$^{-9}$ M$_\odot$ yr$^{-1}$. We estimate that a symmetric bipolar jet would 
transport at the low and intermediate velocities probed by rotation measurements, 
an angular momentum flux of $\dot{L}_j \sim$ 2.9  $\pm$ 1.5 10$^{-6}$  M$_\odot$ yr$^{-1}$ AU km s$^{-1}$.
We discuss implications of these findings for jet launch theories.}
{
The derived  properties of the DG Tau jet  
are demonstrated to be consistent with magneto-centrifugal theory. However,  
non-stationary modelling is required in order to explain all of the features revealed 
at high resolution.
}


\titlerunning{Physical properties of the jet from DG Tau}
\authorrunning{L. Maurri et al.}
   \maketitle

\section{Introduction} \label{intro} 

Herbig-Haro (HH) jets emanating from young stars have been widely studied
in the recent past (\citealt{Bally07}, \citealt{Ray07}). Jets are
believed to regulate important processes, such as the extraction of 
the angular momentum in excess from the star-disk system and the dispersion of the parent
cloud. The phenomenon, however, is far from completely understood,
and  open questions still remain.

For example, the elegant magneto-centrifugal theory behind the proposed
models for the jet launch (e.g., \citealt{Shu00}, \citealt{Ferreira06}, 
\citealt{Pudritz07}, \citealt{Edwards09}) still lacks a strong observational confirmation. 
In this scenario, particles that have been lifted from the disk are then accelerated and collimated by the combined action of centrifugal and magnetic forces along hour-glass shaped magnetic surfaces anchored to the disk. 
This process takes place within the first few AU from the central star, a region that is not directly observable even for the closest jet-disk systems. 
Another important open issue is the nature of the gas excitation. 
The observed jet emission is generally attributed to the presence of shocks that heat the gas locally (\citealt{Hartmann89}, \citealt{Hartigan94},
\citealt{Hartigan95}, and \citealt{Bacciotti99}). However, other heating mechanisms may also be in operation, such as ambipolar diffusion (\citealt{Safier93}; \citealt{Garcia01}) or turbulent dissipation in a viscous mixing layer \citep{Raymond94}.

To clarify these issues it is important to rely on high resolution facilities, such as the Hubble Space Telescope (HST), to examine young stars for which the immediate stellar environment is not opaque, i.e.
classical T Tauri stars (CTTS). 
In 1999, we used the Hubble Space Telescope Imaging Spectrograph (HST/STIS) to observe 
the jet from the CTTS DG Tau at optical wavelengths. The angular resolution of 0.$''$1 corresponds to a spatial scale of 14 AU at the distance of the Taurus cloud (140 pc). Although this scale is much greater than that of the magneto-centrifugal engine (a few AU), we can expect to see an imprint of the jet launching conditions. 

The DG Tau jet (HH~158) was one of the first HH jets
discovered \citep{Mundt83} and, because of its brightness, proximity, 
and structure, it is still one of the most well-studied stellar jets. The blue-shifted lobe of the bipolar jet 
is inclined by about 38$^{\circ}$ to the line of sight 
\citep{Eisloffel98}. 
The flow presents a diverging geometry and
appears to blow a sequence of `bubbles' that terminate in luminous 
bow-like features.  In particular, a large bright 
bow-like structure  was
imaged at 2\farcs7 arcseconds from the source in 1997  
by \cite{Lavalley97} 
(labelled B1 in their nomenclature),  
but other knots are also seen farther away \citep{Eisloffel98, McGroarty07}. 
Subsequently, the DG Tau jet has 
been studied at near
infrared wavelengths (e.g., \citealt{Takami02}, \citealt{Pyo03}, and 
\citealt{Agra-Amboage11}), and most recently in the X-ray domain 
\citep{Guedel08}.

A high resolution study of the DG Tau jet morphology and kinematics was conducted based on HST/STIS observations. In 1999, our team obtained a valuable HST/STIS dataset of the DG Tau jet, consisting of seven long-slit spectra. The slit was placed along the jet and stepped by 0\farcs07 across the jet width, to build a three-dimensional datacube (i.e. two spatial dimensions and one spectral dimension).  
Firstly, \citet{Bacciotti00} presented examples of high spatial
resolution velocity-channel maps of the jet, within the first 2$\arcsec$
from the source. These maps outline well-defined 
features in the flow, as well as
an onion-like kinematic structure in which the low velocity gas is
less collimated than that at higher velocities. Subsequently, 
\citet{Bacciotti02b} 
described how these data provide possible indications for rotation
of the jet about its symmetry axis close to the base of the flow. 
Further signatures of jet rotation 
from complementary HST/STIS observations were presented in 
\citep{Coffey04, Coffey07}. 
These rotation results supported the magneto-centrifugal jet launch
scenario (e.g., \citealt{Ferreira06}, \citealt{Pudritz07}) and, for the first
time, tested the idea that jets can extract angular momentum from the disk,
in order to permit accretion onto the star at the observed rate.
 
In the present study, we continue to exploit the 1999 HST/STIS 
dataset, in order to achieve a detailed parameterisation of the jet plasma physics. 
Previous studies of the gas conditions include: \citet{Lavalley-Fouquet00}; 
\citet{Bacciotti02a} (preliminary analysis of HST/STIS 1999 data); 
and \citet{Coffey08}. These studies relied on the so-called BE-technique \citep{Bacciotti99}, and yet give contradictory reports regarding the correlation between gas excitation and gas velocity. The present study aims to clarify the issue, while providing  high resolution maps of the gas physics  in three dimensions (two spatial and one in velocity). No jet plasma study to-date (\citet{Bacciotti02a}, \citet{Melnikov08, Melnikov09}, \citet{Coffey08} 
\citet{Hartigan07}) has been in a position to present the combination of high resolution with all three dimensions. 

The most important outcome of this study is the determination of the 
{\em total hydrogen density}, resolved in space and velocity. This is a
fundamental parameter for the characterisation of the jet dynamics, allowing 
estimates of the mass outflow rate ($\dot{M}_j$)
and of the angular momentum flux ($\dot{L}_j$).

The paper is organised as follows. The observations and 
the diagnostic techniques are presented in section
~\ref{observ}. The position-velocity (PV) diagrams of the emission lines and their ratios 
are illustrated in section  
~\ref{resu_emiss}. The results of the spectral diagnostic analysis 
are given in section~\ref{resu_phys}, and 
their implications for the dynamics of the system 
are discussed in section~\ref{discuss}. 
Finally, section~\ref{concl} summarises our conclusions.


\section{Observations and method of analysis} \label{observ}

\subsection{Observations and data reduction}

As described in \citet{Bacciotti00}, seven optical spectra of DG~Tau and
its jet were taken with HST/STIS in January 1999 (Proposal ID.
GO 7311). The slit was placed parallel to the jet axis (P.A.
$\sim$226$^{\circ}$), and stepped by 0\farcs07 in the transverse direction thus
covering a total jet width of $\sim$\,0\farcs5. The spectra (labelled
S1, S2, ... S7, from south-east to north-west) constitute a 3-D data-cube, comprising two spatial dimensions and one spectral dimension. 
To observe strong jet tracers such as 
[\ion{O}{i}]\,$\lambda\lambda$6300,6363,
[\ion{N}{ii}]\,$\lambda\lambda$6548,6583,
[\ion{S}{ii}]\,$\lambda\lambda$6716,6731, 
and H$\alpha$, the G750M grating was used, covering a wavelength band of  
652\,\AA\ centred at 6581\,\AA.
The stronger component of the OI doublet, [\ion{O}{i}]$\lambda$6300, is 
partially blue-shifted off the detector. However, since this  doublet is emitted 
in a fixed ratio of 3:1, we use the [\ion{O}{i}]$\lambda$6363 line in its place. 
At the chosen wavelengths, the angular resolution of HST is
0\farcs1, with two-pixel sampling. 
The slit aperture is 52$\times$0.1 arcsec$^2$.
and the spectral sampling 0.554\,\AA\,pixel$^{\rm -1}$.  
The effective velocity resolution is $\sim$ 50 km s$^{-1}$ for extended sources. 
Standard data reduction was carried out by the HST/STIS
pipeline. IRAF tasks were used to remove the effects of bad pixels and cosmic rays, to conduct continuum subtraction, and to convert to a velocity scale. 
Velocities have been corrected for the heliocentric velocity of the star, v$_{\star, hel} \sim$ +17.0\,km\,s$^{\rm -1}$, as derived from a Gaussian fit to the LiI\,$\lambda$6707 photospheric
absorption line in the central slit position. A velocity resampling 
was applied to achieve the same dispersion in all lines (24.67 
km s$^{-1}$ per pixel), to ensure the utmost accuracy in the line ratios.

\subsection{Application of the BE diagnostic technique}
\label{BE}

The BE-technique is a method of obtaining information on the gas physics by comparing ratios of the gas emission lines (\citealt{BE99}, \citealt{Podio06}). 
The technique relies on the fact that, in low excitation conditions and far
from strong sources of ionising radiation, sulphur is ionised only once, and the ionisation state of oxygen and nitrogen is dominated by 
charge-exchange with hydrogen. Under these conditions, the ratios between the 
forbidden emission 
lines emitted by S$^+$, O, and N$^+$ 
are a known function of electron density, $n_e$, ionisation fraction, $x_e$ 
(where $x_e$ = $n_e$ / $n_H$), 
and electron temperature, $T_e$. 

With the lines in our dataset,
$n_e$ can be calculated from the ratio [\ion{S}{ii}]$\lambda6731$
/[\ion{S}{ii}]$\lambda6716$ (hereafter [\ion{S}{ii}]31/16).
Then, using $n_e$, a 
dedicated numerical code (see \citealt{Melnikov08})
evaluates the ratios
[\ion{N}{ii}]($\lambda$6583+$\lambda$6548)/[\ion{O}{i}]($\lambda$6300+$\lambda$6363) 
and [\ion{O}{i}]($\lambda$6300+$\lambda$6363)/[\ion{S}{ii}]($\lambda$6716+$\lambda$6731)
(hereafter [\ion{N}{ii}]/[\ion{O}{i}] and [\ion{O}{i}]/[\ion{S}{ii}], respectively) against
a grid of $x_e$ and $T_e$ values.
The best fit gives the ionisation (and
temperature) of the emitting gas,
leading ultimately to an estimate of the total hydrogran density, $n_H$, a fundamental parameter in jet dynamics.

The procedure is independent 
of the assumed heating mechanism and its simplicity of application
allows speedy investigation of large datasets.
Excitation conditions are assumed to remain constant along the line of sight,
which results in smoothing the gradients  of the quantities \citep{DeColle10}. However, in the present case the problem is mitigated by 
the velocity resolution, which naturally sorts the different jet layers. 
A further limitation is that in 
spatially unresolved shock waves, $T_e$ varies rapidly over the 
line emission region, while the evolution of $n_e$ and $x_e$ 
is slow. Therefore, as discussed in \citet{BE99}, 
the observed line ratios, averaged over the resolution
element and along the line of sight, can only give a rough indication of the 
local excitation temperature. 
In order to illustrate the uncertainties of the BE-technique in the 
case of unresolved shocks, a determination of 
$x_e$ has been attempted from the grid of shock models 
of \citet{Hartigan94}, assuming that the ionisation in the gas 
is produced {\em locally} in each resolution element by a shock. 
To this aim, we used the grid of shock models by \citet{Hartigan94}, finding 
values of $x_e$ lower than those derived with the BE-technique by about 30\%.

Finally, extinction is not taken into account, as 
this should be determined locally around the jet base 
using emission lines across a broader wavelength range. 
However, the lines used in the BE technique are close in 
wavelength, and typically in the case of CTTS, the uncertainty 
introduced by not accounting for extinction is found to be lower 
than the error due to noise \citep{Podio06}. We also note that 
to increase the number of positions where an indication of the plasma conditions 
can be given, in regions where one of the emission line intensities falls 
below 3$\sigma$, the 3$\sigma$ value is used in order to obtain an upper/lower limit for the ratio. 
In this case the results of the diagnostics are given in terms of upper or lower limits.

In this work we use an updated version of the technique with
respect to recent papers
\citep{Coffey08, Melnikov09}.
As in \cite{Podio11}, we use values for the 
collision strengths derived 
from the results of \cite{Keenan96} for S$^+$, and of 
\cite{Hudson05} for N$^+$. 
For neutral oxygen, we use 
the values of the collisional coefficients reported in \cite{Berrington81}, which integrate 
the compilation by \cite{Mendoza83}. The interpolation of   
these coefficients gives 
better values of the collision strengths over a wider
temperature range than in previous studies. 
Elemental abundances are taken from \cite{Asplund05}.

In addition, where the plasma density is higher than the high  density limit 
for the [\ion{S}{ii}] ratio, thus preventing derivation of electron density, 
we use an extension of the diagnostic code 
which relies instead on the [\ion{N}{ii}]/[\ion{O}{i}] and[\ion{O}{i}]/[\ion{S}{ii}] line ratios to find $n_e$ and $x_e$. 
This extension assumes a value of T$_e$ derived for neighbouring points where
the standard BE technique can be safely applied (see sect.\ref{resu_phys}). 
In this case the uncertainty in the 
determination of $n_e$  and $x_e$ is estimated to be 
of about 25\% and 15\%, respectively, for  
variations of the assumed temperature of 30\%, with both quantities 
decreasing for increasing $T_e$. When [\ion{N}{ii}] is below the 3$\sigma $ threshold, 
the inferred values of $x_e$ and $n_e$ are upper limits.

\section{Results: 3-D kinematic structure and line ratios} \label{resu_emiss}

We present our data and results as PV maps of the emission lines, of the line ratios, and of the derived plasma parameters. In each figure the seven PV maps obtained from the stepped slit positions  cover a combined field of view of $\sim$ 5$\arcsec \times$ 0\farcs5. The dashed lines indicate the position of emission peaks which were identified in previous studies: A2 at 0\farcs75 and A1 at 1\farcs45 in this dataset, \cite{Bacciotti00}; B1 at 2\farcs7 \cite{Lavalley97} in 1998, seen in this dataset at 3\farcs8 (B1 is the same feature as the X-ray feature at 6$''$ identified in 2010 \cite{Guedel11}, based on proper motion of 0$\farcs$275 yr$^{-1}$ \citep{Pyo03}); and a secondary peak, B0, at 3\farcs3, identified in the channel maps at high velocity of \citet{Lavalley-Fouquet00}. 

\subsection{Position-velocity diagrams of the surface brightness} 
\label{spectra}

\begin{figure*}[!hptb]
 \resizebox{\hsize}{10.8cm}{\includegraphics[angle=90]
{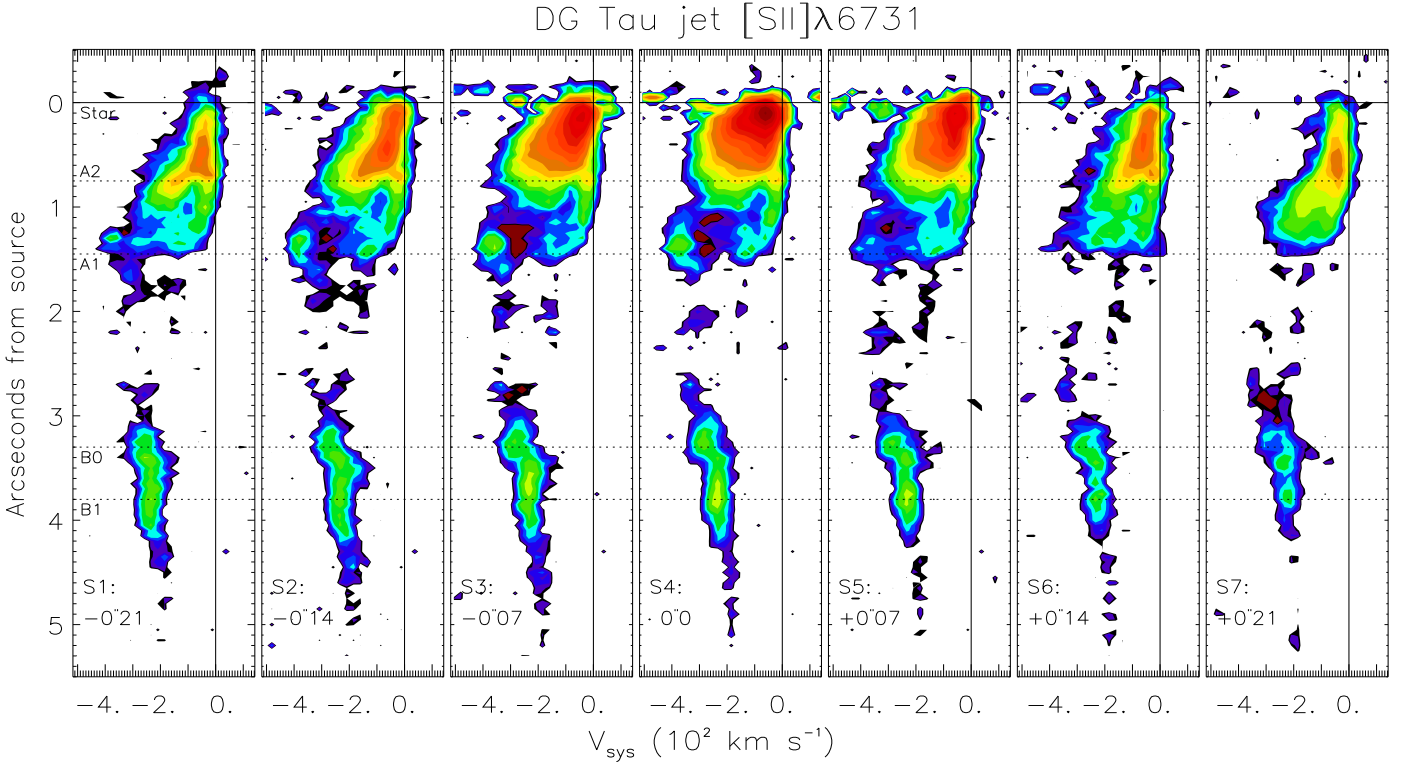}}
  \caption{Continuum-subtracted HST/STIS position-velocity (PV) plots 
of the jet from DG Tau, in [\ion{S}{ii}]$\lambda$6731 emission in
slit positions S1 to S7 (south-east to north-west). Contours are from
1.1~10$^{-15}$\,erg\,s$^{-1}$\,arcsec$^{-2}$\,cm$^{-2}$\,\AA$^{-1}$
(3 $\sigma$), with a ratio of 2$^{2/5}$. The solid lines
mark the position of the star and zero velocity, 
while dashed lines mark the positions of
identified features in images of this flow (see text). 
}
  \label{SII31spec7}
\end{figure*}

\begin{figure*}[!hptb]
 \resizebox{\hsize}{10.8cm}{\includegraphics[
angle=90]{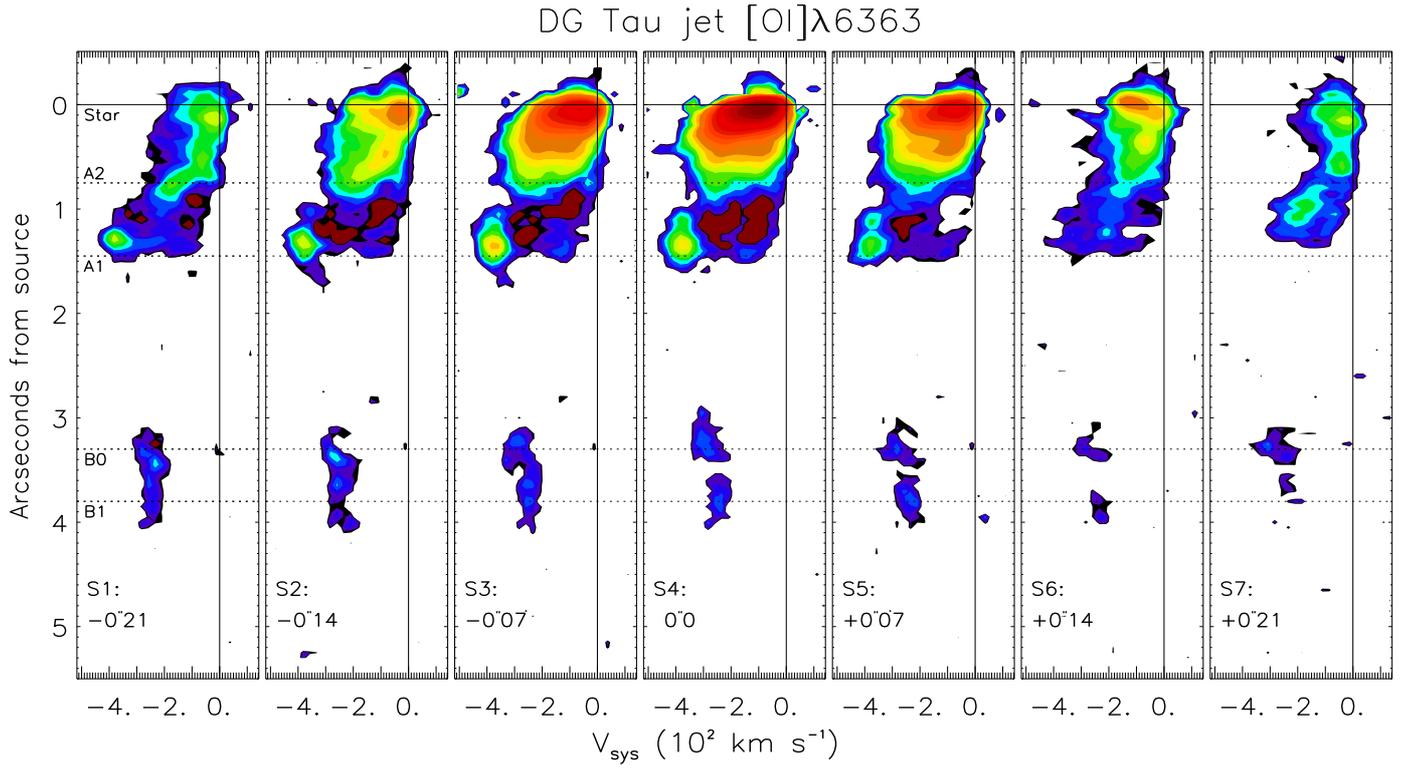}}
 \caption{Same as Fig.~\ref{SII31spec7},
but for [\ion{O}{ii}]$\lambda$6363.
The stronger doublet component, [\ion{O}{i}]$\lambda$6300, is blue-shifted off the detector. However, since this  doublet is emitted in a fixed ratio of 3:1, we use the [\ion{O}{i}]$\lambda$6363 line in its place. 
} 
  \label{OI63spec7}
\end{figure*}

\begin{figure*}[!hptb]
\resizebox{\hsize}{10.8cm}{\includegraphics[angle=90]
{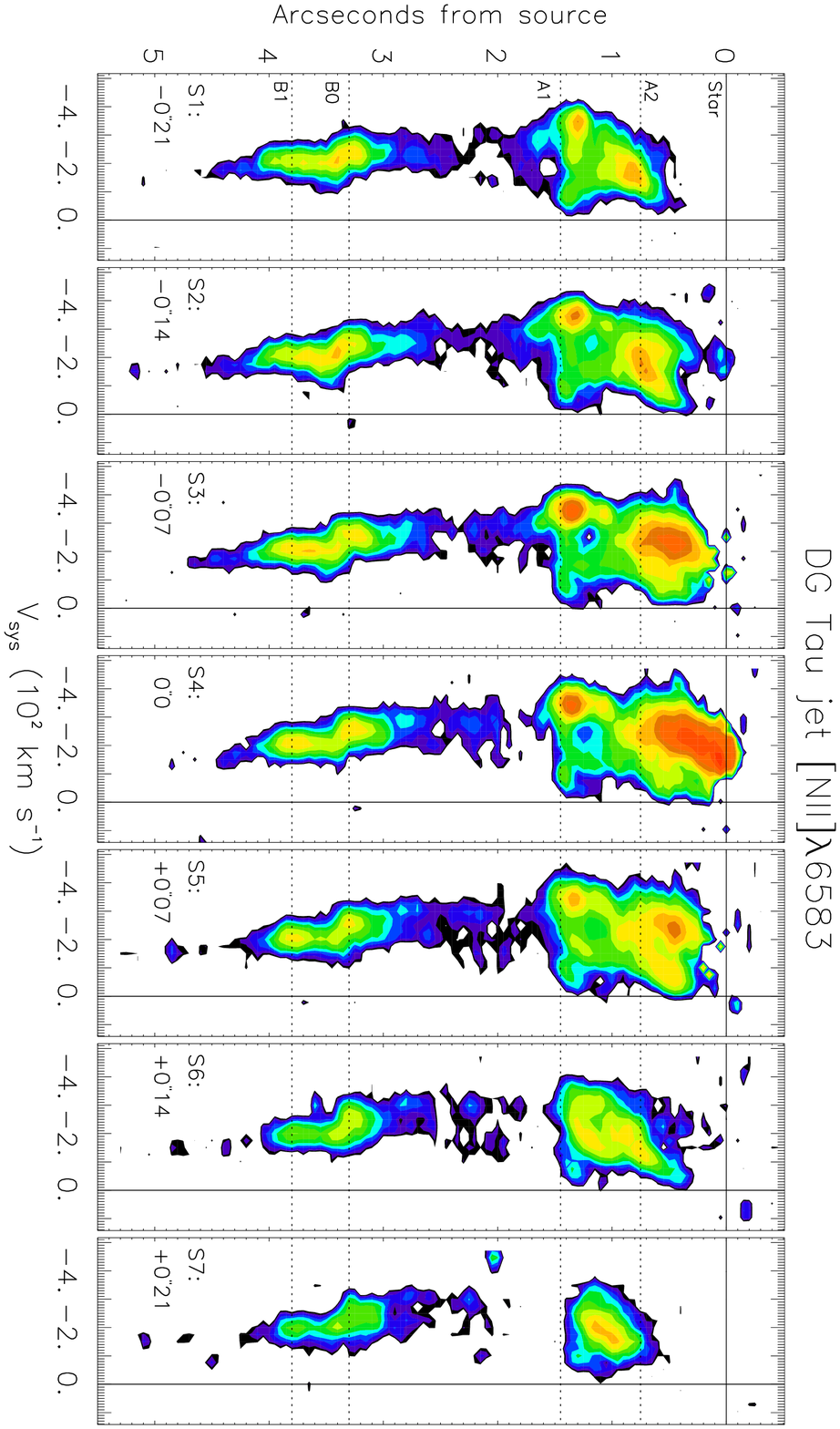}}
  \caption{Same as Fig.~\ref{SII31spec7},
but for [\ion{N}{ii}]$\lambda$6583.}
  \label{NII85spec7}
\end{figure*}

Figs.~\ref{SII31spec7},~\ref{OI63spec7},
and~\ref{NII85spec7} show the 
[\ion{S}{ii}]$\lambda$6731, [\ion{O}{i}]$\lambda$6363 and
[\ion{N}{ii}]$\lambda$6583 emission lines respectively, in three dimensions: along the jet, across the jet, and in velocity space. 


At least three velocity components can be identified in the jet up to location A1:  
a low velocity component, between about -150  and +20,
more evident in [\ion{S}{ii}] and [\ion{O}{i}], and in the lateral positions;  
a medium velocity component between -300 and -150 km s$^{-1}$, more evident in [\ion{N}{ii}], 
characterised by an emission peak at about 0\farcs45 and a slow 
acceleration; and a high velocity component between about -300 and -400 km s$^{-1}$, with  
an emission spot at 1\farcs4, hereafter A1HV, brighter in slits 1, 2, 3 and 4, 
already identified as a knot 
in \citet{Lavalley-Fouquet00}  at 0\farcs93 from the source.
A bifurcation of the emission is evident  
between the low and the medium velocity components
(e.g. slit 1 to 6 in [\ion{S}{ii}] at A2, and slit 5 in  [\ion{N}{ii}] at 0\farcs45). 
A steep gradient towards lower velocities, of $\sim$ 30 km s$^{-1}$, 
is apparent about 0\farcs2 downstream from A1HV in [\ion{N}{ii}], and less evidently in [\ion{S}{ii}].

Futher along the jet, between A1 and $\sim3''$, emission is very low in [\ion{O}{i}] and [\ion{S}{ii}], 
while the [\ion{N}{ii}] line is stronger, mainly at medium 
velocities. This region corresponds to a faint stripe connecting A1 and 
B1 in the images of \citet{Dougados00}. 
The PV plots are very difficult to read here, but 
there are indications of material flowing at two different velocities 
(about -150 and -320 km s$^{-1}$), and marginal indications of a localised gradient in velocity
(of about 30 km s$^{-1}$) at 2\farcs8. 

Beyond 3\arcsec, the system of slits intercepts the large ($\sim$ 2$''$) 
bow-like feature of which B0 and B1 form a part. Here
all lines are detected, with [\ion{N}{ii}] being the strongest, but 
only at velocities between -150 and -350 km s$^{-1}$, and 
brighter in slits S1 to S4. 
Another gradient towards lower velocities, of $\sim$ 70 km s$^{-1}$, is seen at 3\farcs45
between the emission peaks at B0 and B1, and a less evident 
one at 4\farcs1, of $\sim$ 30 km s$^{-1}$, downstream of knot B1.
Interestingly, these apparent abrupt reductions in velocity occur immediately downsteam of a peak in emission. This is expected in radiative shocks, in which the discontinuity in velocity is at the front,
while the optical emission arises behind the shock front on scales resolvable by HST \citep{Hartigan94}.


\subsection{Position-velocity diagrams of the line ratios}

\begin{figure*}[!tphb]
\resizebox{\hsize}{10.8 cm}{\includegraphics[angle=90]
{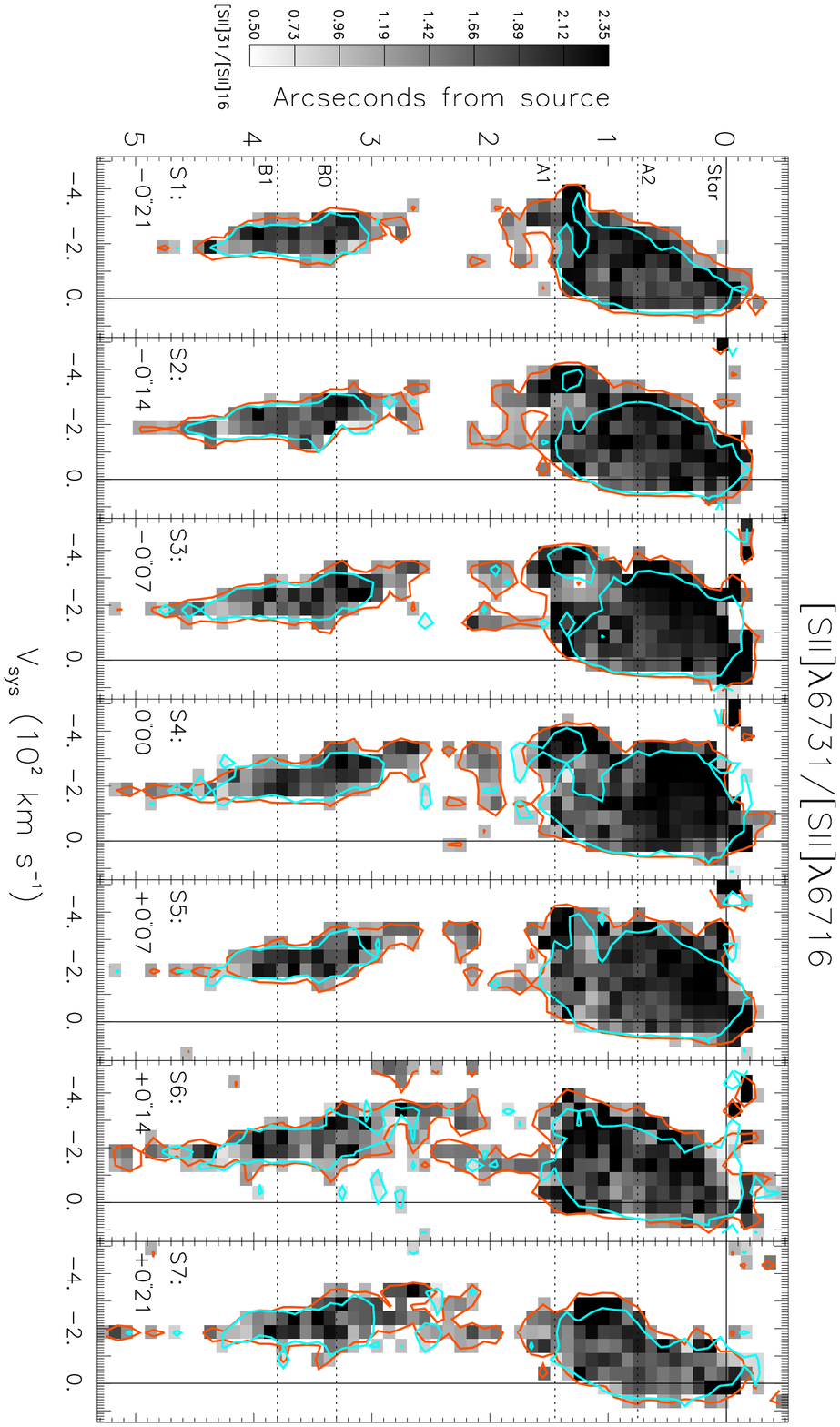}}
  \caption{
PV plots of the [\ion{S}{ii}]$\lambda$6731/[\ion{S}{ii}]$\lambda$6716 
line ratio in linear greyscale.
Cyan and orange contours indicate the [\ion{S}{ii}]$\lambda$6716 and
[\ion{S}{ii}]$\lambda$6731 emission at $3\sigma$, respectively. 
Where only one of the two lines is above the $3\sigma$ threshold 
the upper/lower limit of the line ratio is reported.
}
  \label{s31_s16}
\end{figure*}
\begin{figure*}[!tphb]
\resizebox{\hsize}{10.8 cm}{\includegraphics[angle=90]
{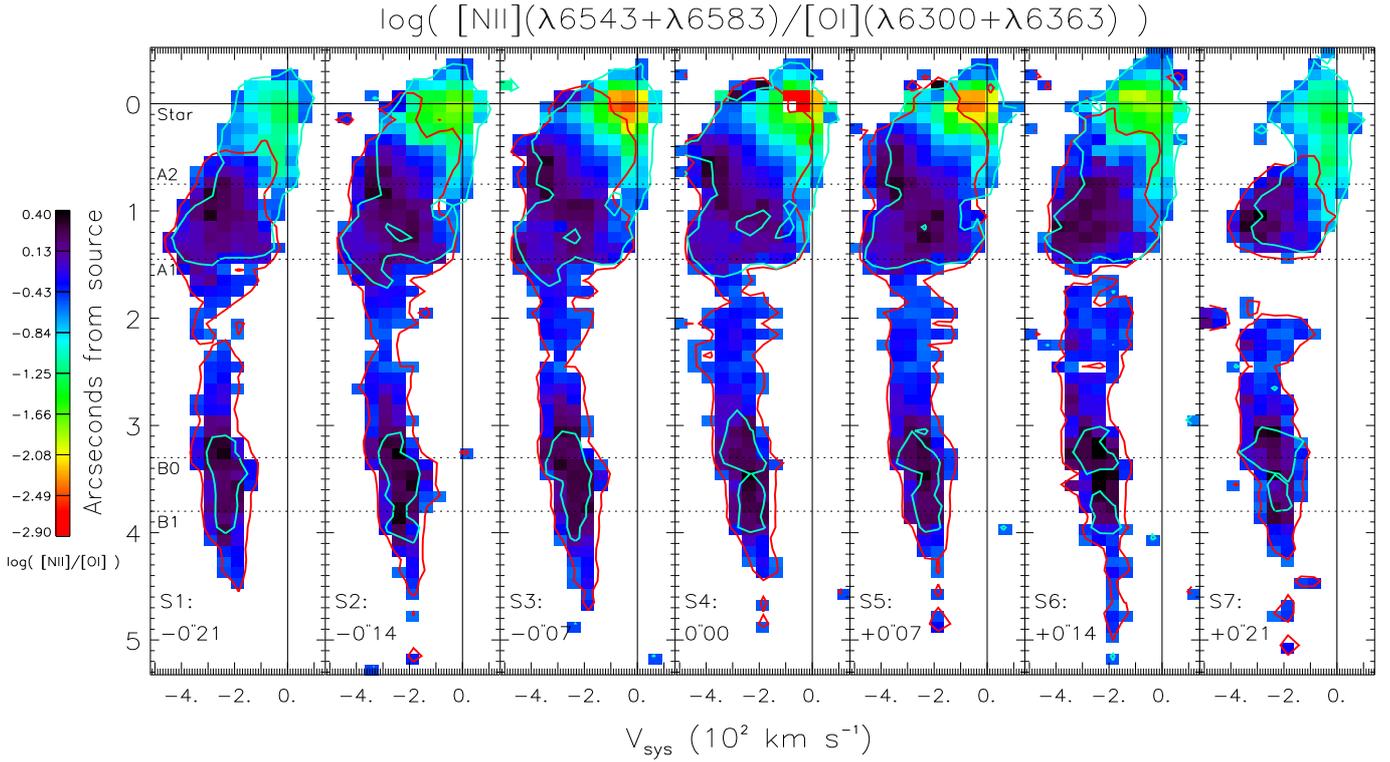}}
  \caption{
Same as Fig.~\ref{s31_s16} 
for the logarithm of the
[\ion{N}{ii}]($\lambda$6583+$\lambda$6548)/
[\ion{O}{i}]($\lambda$6300+$\lambda$6363) line ratio. 
The superposed {\bf cyan } 
and red contours indicate the [\ion{O}{i}]$\lambda$6363 and the
[\ion{N}{ii}]$\lambda$6583 emission at $3\sigma$
([\ion{O}{i}]$\lambda$6300 = 3 [OII]$\lambda$6363 and 
[\ion{N}{ii}]$\lambda$6548 = 1/3 [\ion{N}{ii}]$\lambda$6583 
is assumed everywhere, see text). 
}
  \label{nii_oi}
\end{figure*}
\begin{figure*}[!tphb]
\resizebox{\hsize}{10.8 cm}{\includegraphics[angle=90]
{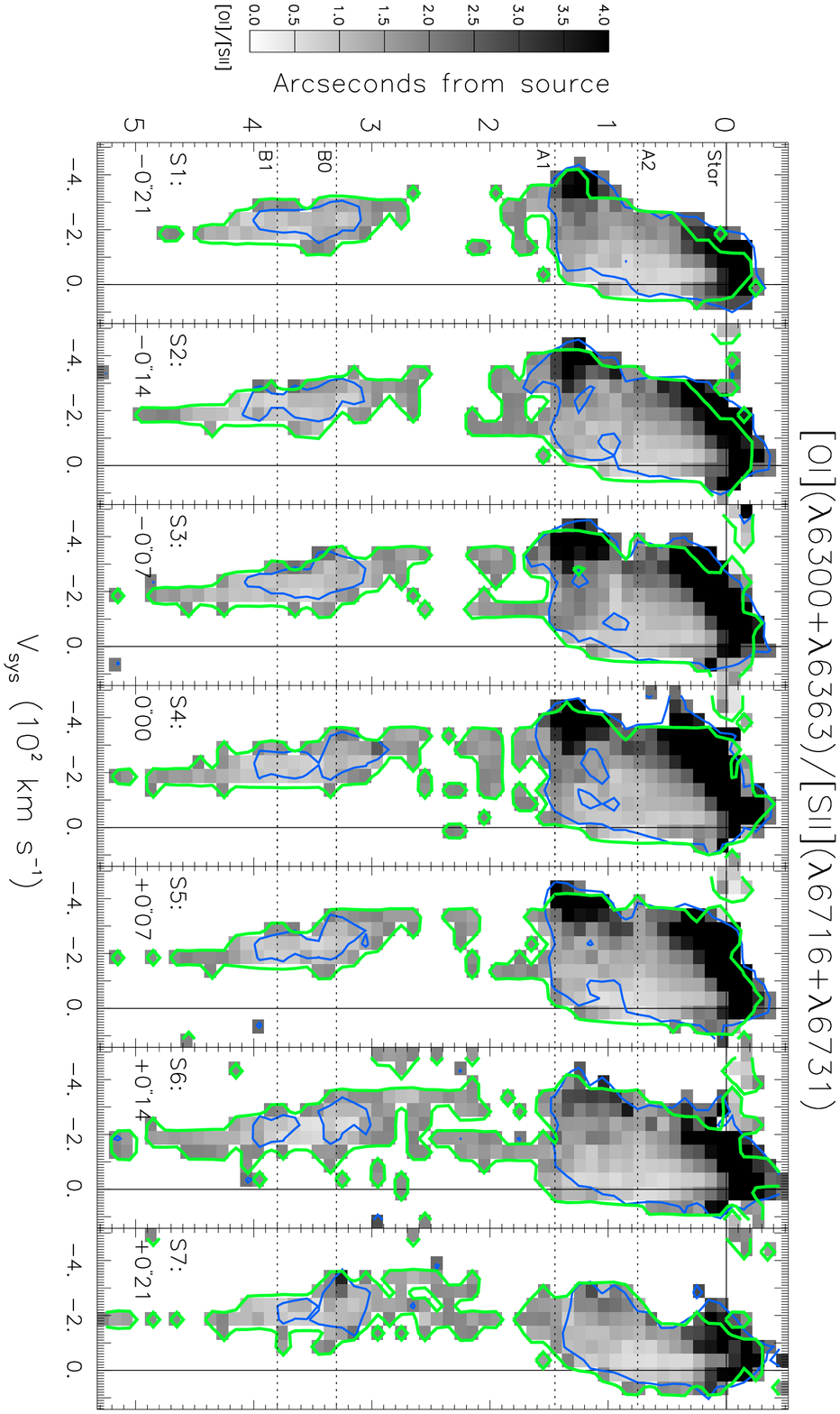}}
  \caption{
Same as Fig.~\ref{s31_s16} for
[\ion{O}{i}]($\lambda$6300+$\lambda$6363)/[\ion{S}{ii}]($\lambda$6716+$\lambda$6731) line ratio.
Blue and green contours indicate 
[\ion{O}{i}]$\lambda$6363 and the [\ion{S}{ii}]6731+6716 emission at $3\sigma$ respectively.
}
  \label{oi_sii}
\end{figure*}

Figures \ref{s31_s16},~\ref{nii_oi}, and~\ref{oi_sii} show 
emission line ratio PV plots for [\ion{S}{ii}]31/16, [\ion{N}{ii}]/[\ion{O}{i}] and  [\ion{O}{i}]/[\ion{S}{ii}], respectively. 
Plots are produced after two-pixel binning in both spatial and spectral dimensions, to reflect resolution (i.e. 0\farcs1 and 50 km s$^{-1}$). 
A 3$\sigma$ contour ($\sim$2.2 10 $^{-15}$\,erg\,s$^{-1}$\,arcsec$^{-2}$\,
cm$^{-2}$\,\AA$^{-1}$ after binning) for each emission line is overlaid. 

The [\ion{S}{ii}] ratio is in many regions 
at the high density limit of 2.35 (HDL), beyond which  
the ratio is no longer a valid diagnostic tool. In these regions, encircled by green contours,  
the electron density is $\geq$  2~10$^4$ cm$^{-3}$, i.e. high 
when compared to published values at large distances along many jets 
of typically 10$^3$ cm$^{-3}$.

The [\ion{N}{ii}]/[\ion{O}{i}] ratio is
a good indicator of the hydrogen ionisation fraction 
since it increases monotonically with it,
and is nearly independent of $n_e$ and $T_e$, as long as 
the electron density is below the [\ion{N}{ii}] critical electron 
density of $n_e < 10^5$ cm$^{-3}$. 
The value of this ratio is low near the star, but smoothly increases   
with distance and velocity. It reaches higher values at high speeds approaching A2, 
and at medium velocities between A2 and A1.
At A1HV, both the [\ion{O}{i}] and [\ion{N}{ii}] lines 
are intense giving a moderate ratio. 
In the A1-B0 ridge the reported value of $\sim$0.4 is 
a lower limit, as [\ion{O}{i}] has been set to 3$\sigma$. Meanwhile, 
high ratio values return in the B0 - B1 region. 

Finally, the [\ion{O}{i}]/[\ion{S}{ii}] line ratio 
 depends on both $n_e$ and $T_e$, 
and weakly on $x_e$ through [\ion{O}{i}]. The ratio
shows moderate values
almost everywhere except for the shoulder 
between the star and A2 at progressively higher speeds, 
and at A1HV. 
This 'shoulder' appears not to correspond to any of the kinematic components
identified above.   

\begin{figure*}[!phbt]
 \resizebox{\hsize}{10.8cm}{\includegraphics[angle=90]{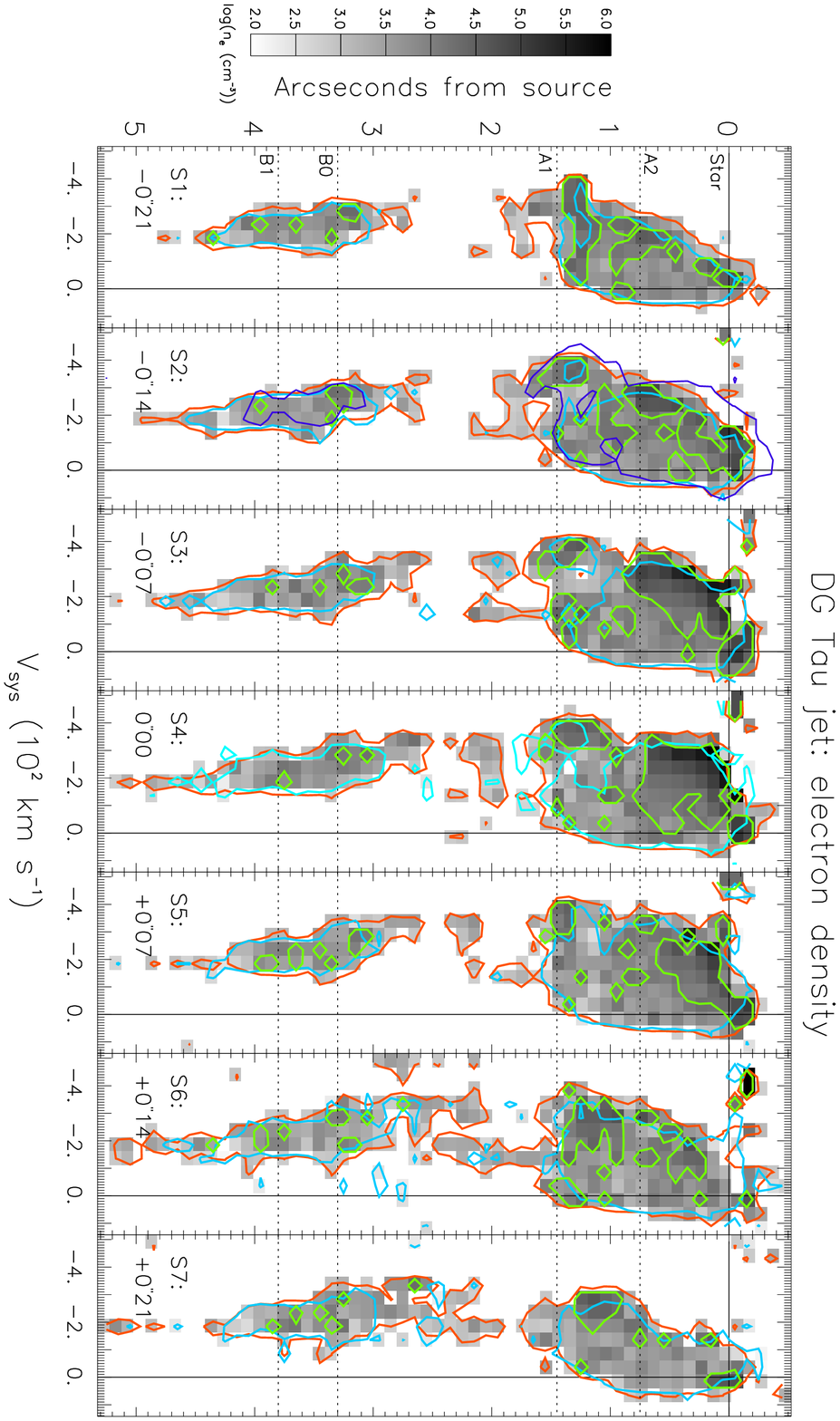}}
  \vspace{-0.15cm}
  \caption{
PV plots of the logarithm of the electron
density $n_e$. Contours indicate regions inside which: 
{\em cyan} - [\ion{S}{ii}]$\lambda6716 \geq ~3\sigma$;
{\em orange} - [\ion{S}{ii}]$\lambda6731 \geq ~3\sigma$;
{\em green} - [\ion{S}{ii}]31/16 ratio at the high density limit.
Where only [\ion{S}{ii}]$\lambda$6731 ([\ion{S}{ii}]$\lambda$6731) $\geq ~3\sigma$, 
the derived $n_e$ is a lower (upper) limit. Inside the green contours 
$n_e$ is derived with the modified HDL-BE procedure (cfr. Sect. \ref{BE}).
}
  \label{PVne}
\end{figure*}
\begin{figure*}[!pbht]
\resizebox{\hsize}{10.8cm}{\includegraphics[angle=90]{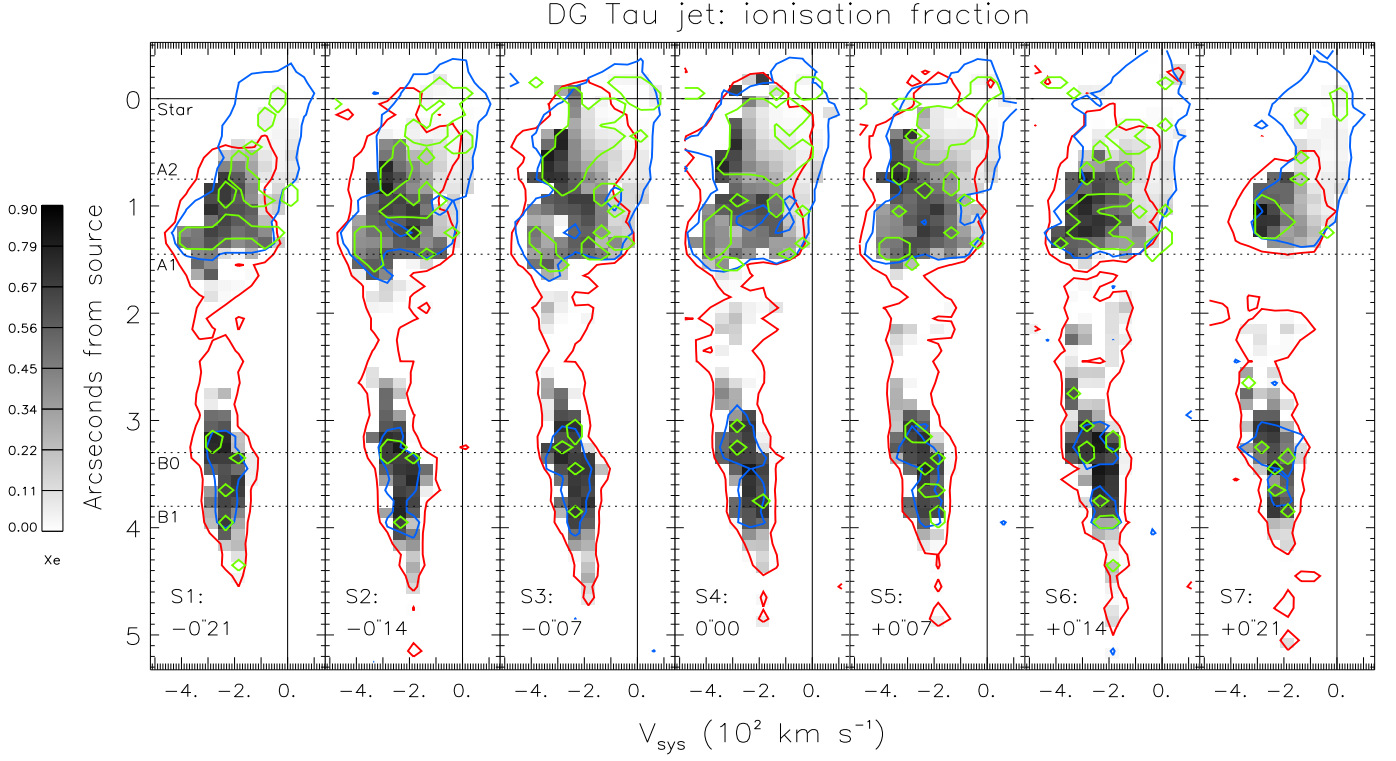}}
  \caption{Position-velocity diagrams of the ionisation fraction 
x$_e$. Contours: 
{\em blue} - [\ion{O}{i}]$\lambda$6363 at $3\sigma$; 
{\em red} - [\ion{N}{ii}]$\lambda$6583 at $3\sigma$; 
{\em green} - high density limit for the [\ion{S}{ii}]31/16 ratio.
Where only [\ion{O}{i}] ([\ion{N}{ii}]) is above $3\sigma$, $x_e$ is 
an upper (lower) limit. Inside the HDL regions $x_e$ is determined
with the HDL-BE procedure (cfr. Sect. \ref{BE}).
}
  \label{PVxe}
\end{figure*}

\begin{figure*}[!pbht]
\resizebox{\hsize}{10.8cm}{\includegraphics[angle=90]{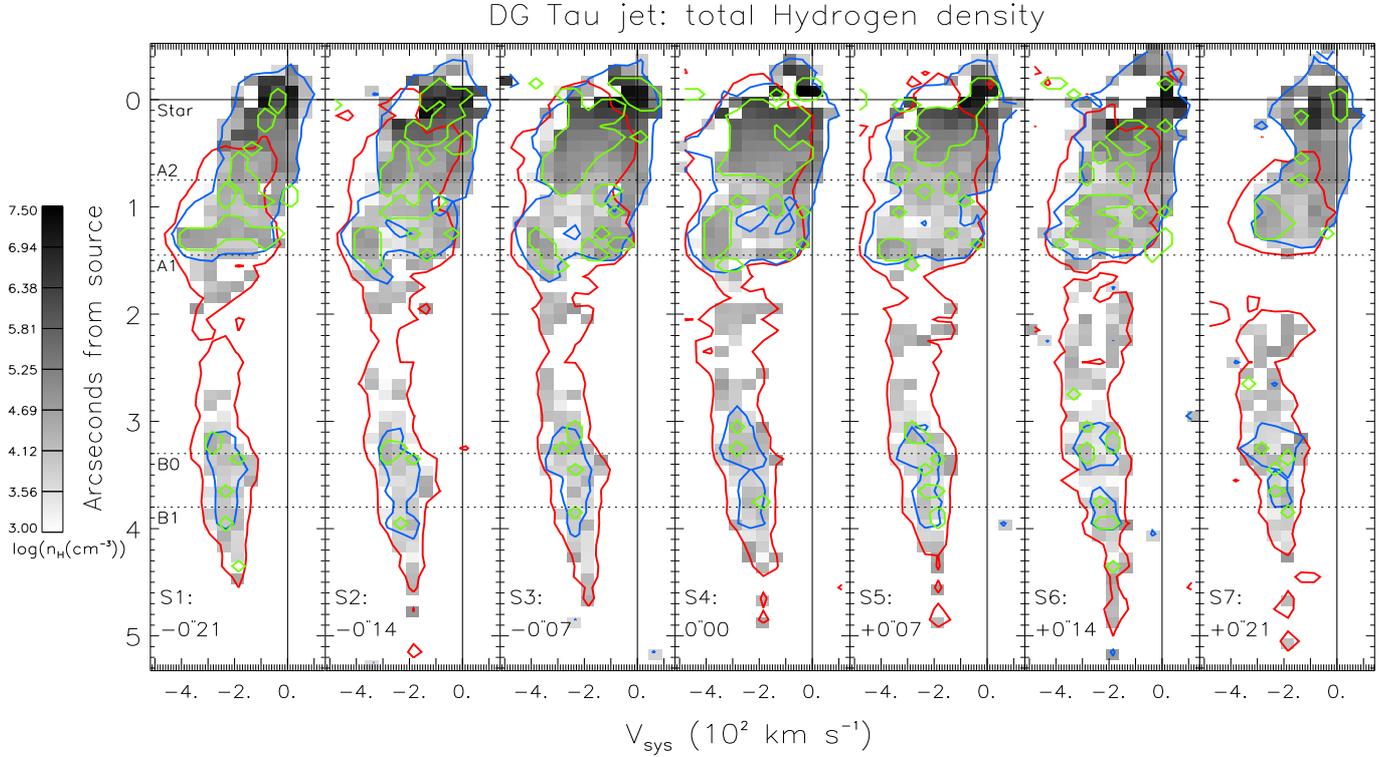}}
  \caption{Position-velocity diagrams of the logarithm of the hydrogen
density derived as n$_H =$ n$_e$/x$_e$. Contour colour coding is as in 
Fig.\ref{PVxe}.
}
  \label{PVnh}
\end{figure*}
\begin{figure*}[!bpt]
\resizebox{\hsize}{11.5cm}{\includegraphics{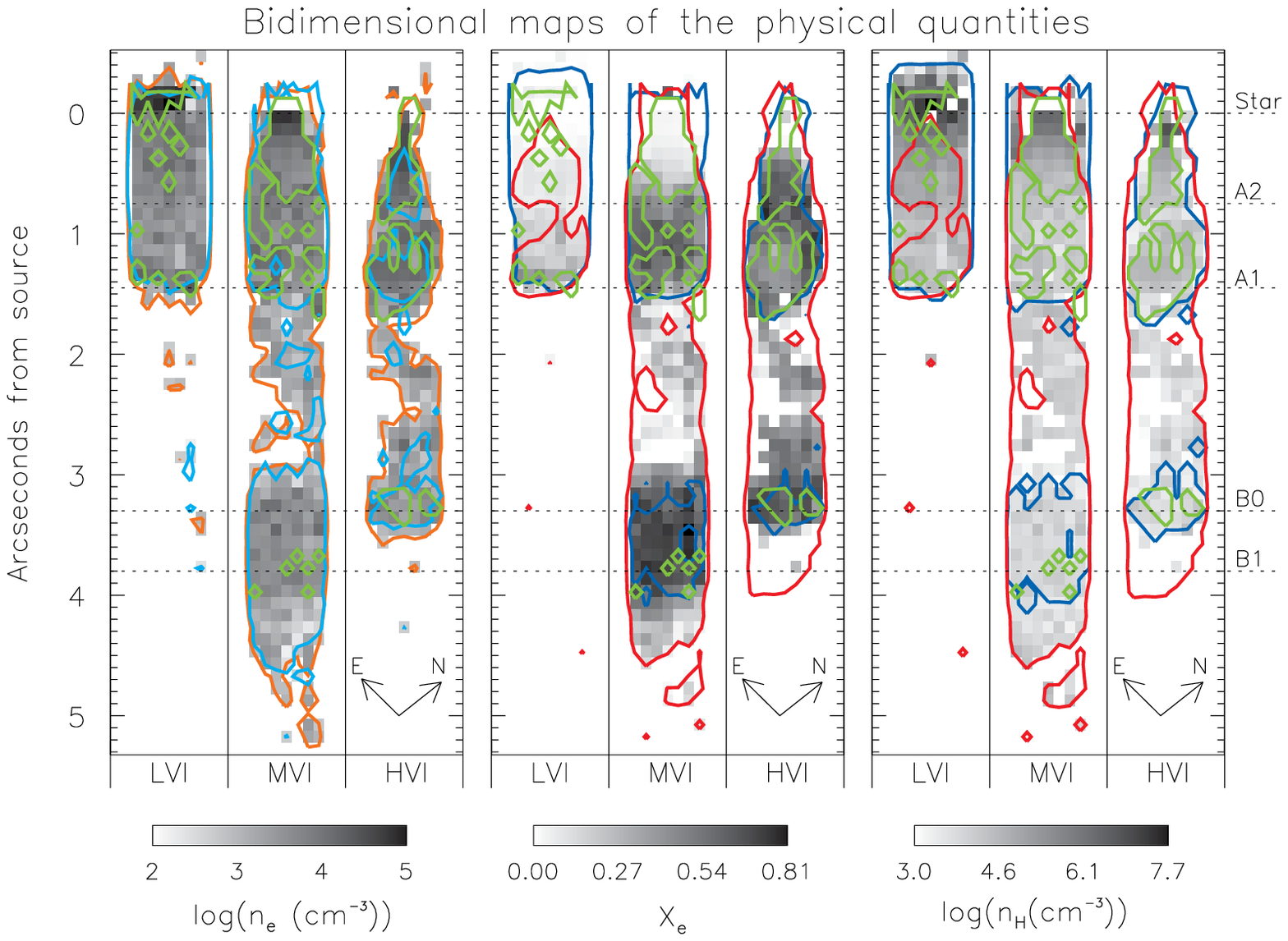}}
   \caption{
2-D velocity channel maps of 
the logarithm of the electron density (left), of
the ionisation fraction (centre)  and of the logarithm of the total density (right).  
Greyscales are linear. 
Low velocity interval (LVI) is defined as -120 to +25 km s$^{-1}$, 
medium velocity interval (MVI) is defined as -270 to -120 km s$^{-1}$ and 
high velocity interval (HVI) is defined as -420 to -270 km s$^{-1}$. 
Contours: 
{\em cyan} - [\ion{S}{ii}]$\lambda$6716 at $3 \sigma$; 
{\em orange} - [\ion{S}{ii}]$\lambda$6731 at $3 \sigma$; 
{\em green } -  region above [\ion{S}{ii}] critical density; 
{\em blue} - [\ion{O}{i}]$\lambda$6363 at $3 \sigma$; 
{\em red} - [\ion{N}{ii}]$\lambda$6583 at $3 \sigma$. 
}
\label{bimapEC1}
 \end{figure*}

\section{Results: physical quantities and jet widths} \label{resu_phys}

The diagnostic analysis was performed in three ways, in order to optimise the output and to facilitate comparisons with the literature. The first preserves all high resolution information in 3-D (i.e. 2-D space and 1-D velocity), and the results are presented as PV plots of $n_e$, $x_e$, and $n_H$ (Fig.~\ref{PVne}, \ref{PVxe}, \ref{PVnh}) giving a global picture of the jet excitation conditions. The results highlight regions of low signal-to-noise where the diagnostics would benefit from a further binning of the input data in space and/or velocity. Therefore, the second way of carrying out the analysis bins the velocity information into three velocity channels 
(Fig.~\ref{bimapEC1}),
defined as low velocity interval (LVI), from -120 to +25 km s$^{-1}$, 
medium velocity interval (MVI), from -270 to -120 km s$^{-1}$ and 
high velocity interval (HVI) from -420 to -270 km s$^{-1}$. 
The third way bins in velocity and jet width giving plasma parameters in 1-D along the jet as it propagates (Figs.~\ref{prof1-Dne}, \ref{prof1-Dxe}, \ref{prof1-Dnh}). 1-D profiles of the line ratios are provided as on-line material, Figs.~\ref{1Ds3116} - \ref{1Do_s}. 
In all cases, in positions in which the [SII] ratio is at 
the high density limit, the modified HDL technique has been applied
(cfr. Sect. \ref{BE}) adopting in the affected regions  T$_e = 10^4 K$ for the LVI, 
T$_e =$ 1 and 2 10$^4$ K for the MVI, upstream and downstream A2, respectively, and  
T$_e =$ 3 10$^4$ K for  the HVI. 
Lastly, the jet width of the various velocity components is estimated, Fig.~\ref{fwhm}.

\subsection{Electron density} 
\label{ne}

\begin{figure}[!hptb]

 \resizebox{\hsize}{!}{\includegraphics{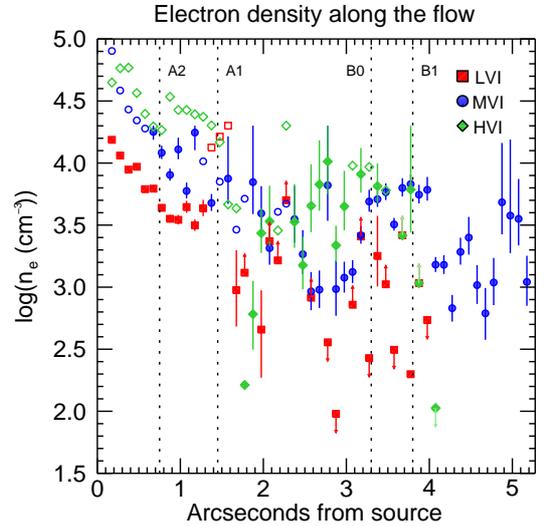}}
       \caption{
1-D profiles of  
log(n$_e$) along the flow in discrete velocity intervals (see Fig. \ref{bimapEC1}).
Error bars (smaller than the symbol size where the signal is strong) 
are determined a posteriori by  
conducting the analysis with input values of the line ratios 
and their uncertainties 
(evaluated from the 3$\sigma$ error on the fluxes).
Empty symbols indicate positions where n$_e$ is above 
the high density limit ofr the [\ion{S}{ii}] diagnostics, 
and the derivation has been made through 
the modified BE-technique described in Sect.~\ref{BE}.
}
         \label{prof1-Dne}
   \end{figure}

The results for $n_e$ are illustrated in Fig.~\ref{PVne}, 
~\ref{bimapEC1} (left panels) and \ref{prof1-Dne}. 
The electron density is higher than 10$^3$ cm$^{-3}$ almost everywhere in the jet (Fig.~\ref{PVne}). The densest portion is near the jet base as far as A1, and for high velocities.
While this trend was previously reported by \citet{Bacciotti00} and \citet{Bacciotti02a}, these studies were limited by the critical density of [\ion{S}{ii}]. Closer inspection via the modified BE-technique 
reveals that $n_e$ is higher in the MVI and HVI  than in the LVI (Fig.~\ref{bimapEC1}), 
with values reaching close to 10$^5$ cm$^{-3}$ near the star. 
Between the star and A2, there is marginal evidence of a separation
between the LVI and the other velocity components (Fig.~\ref{PVne}).
Meanwhile, there is no variation of $n_e$ across the jet, from S1 to S7 in each velocity interval
(Fig.~\ref{bimapEC1}) .

Further along the jet, a slight increase in $n_e$ is noted at A1HV (Fig.~\ref{PVne}), while 
downstream of A1, where the flow is seen only in the MVI and HVI, $n_e$ continues to decrease in the MVI, reaching 10$^3$ cm$^{-3}$ at 2\farcs8, in contrast to the HVI which reaches a maximum at this position. Continuing to the B0-B1 region, $n_e$ increases again in the MVI. Finally, the HVI contribution fades after B0, while in the MVI $n_e$ is detectable well beyond B1. 

\subsection{Ionisation fraction}
\label{xe}

\begin{figure}[!htb]

 \resizebox{\hsize}{!}{\includegraphics{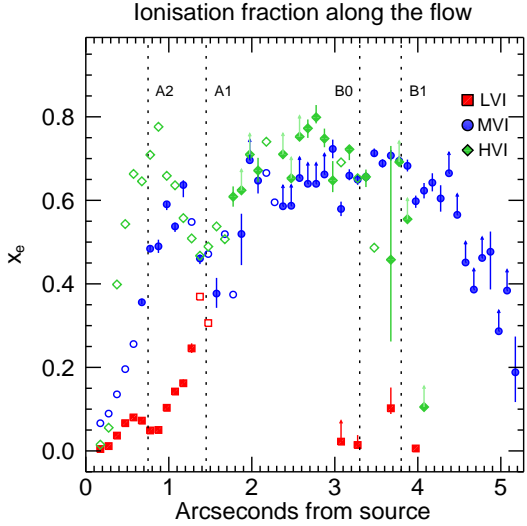}}
      \caption{
Same as Fig.~\ref{prof1-Dne} for x$_e$. 
}
         \label{prof1-Dxe}
   \end{figure}

The results for $x_e$ are illustrated in Fig.~\ref{PVxe}, ~\ref{bimapEC1} (middle panels) 
and ~\ref{prof1-Dxe}.
It is worth noting that the 'missing' $x_e$ data points result from a  
low signal in both [\ion{S}{ii}] lines, or lack of
an unique solution from the code where [\ion{O}{i}]/[\ion{S}{ii}] $>>$1
(i.e. between the star and A2).
However, the trends in those regions can be gleaned from the [\ion{N}{ii}]/[\ion{O}{i}] ratio (Fig.\ref{nii_oi}),
which is almost directly proportional to $x_e$. 

At the base of the jet, $x_e$ is not larger than 0.07 (Fig.~\ref{PVxe}), 
but it increases with distance and velocity up to a high value of over 0.6 
in the region between A2 and A1, peaking at different distances in each velocity interval. 
Before  A1, $x_e$ drops in MVI and HVI, while in LVI it 
increases toward A1 
(Fig.~\ref{PVxe}). 
There is no strong enhancement of $x_e$ at A1HV, 
contrary to what would be expected 
given the strong emission. 
Moving along the jet, we note that between A1 and B0 the determination of $x_e$ is poor,
and given often in terms of lower limits. 
Binning over space and velocity increases the [\ion{O}{i}] signal-to-noise, 
making more evident
the existence of a high ionisation region between 2$''$ and 4\farcs1 in  both the MVI
and the HVI.  In the 1-D profiles $x_e$ shows a plateau in $x_e$ of $\sim$ 0.7 in this region. 
At 3\farcs9 a sharp drop occurs for the MVI, 
possibly coincident with the 30 km s$^{-1}$ velocity gradient (Section~\ref{spectra}). After this point one sees
a more gradual decrease in $x_e$, for the MVI, represented mainly by lower limits (Fig.~\ref{prof1-Dxe}).
Finally, as for $n_e$, no large transverse variations are 
evident in the velocity channel maps of $x_e$ (Fig.~\ref{bimapEC1}).

\subsection{Total hydrogen density} 
\label{nh}

\begin{figure}[!htb]

 \resizebox{\hsize}{!}{\includegraphics{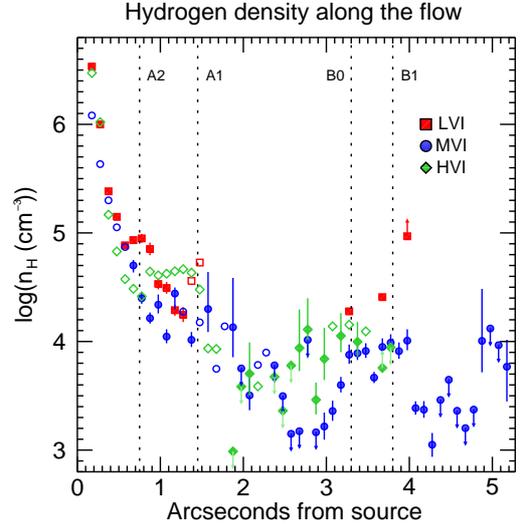}}
  
      \caption{
Same as Fig.~\ref{prof1-Dne} for log(n$_H$). 
}
         \label{prof1-Dnh}
   \end{figure}

The results for $n_H$ are illustrated in Fig.~\ref{PVnh}, \ref{bimapEC1} (right panels) and \ref{prof1-Dnh}.
We stress that our mapping of this quantity is 
more extended than in previous works, thanks to the application of 
the modified BE-technique. Similar to $n_e$, the total hydrogen density 
is at its maximum of 3 10$^6$ cm$^{-3}$ close to the star, 
and then decreases by several orders of magnitude along the flow. 
The range of variation is larger than in n$_e$, 
because of the effect of the increase in ionisation. 
Again, no spatial variation is evident in the transverse direction. 

In contrast to MVI and LVI trends,
moving from A2 to A1 we see a plateau in $n_H$ for HVI, 
giving rise to the high emission of the AHV1 spot. Then $n_H$
drops back again downstream of A1HV (Fig.~\ref{prof1-Dnh}).  
Other $n_H$ increases in HVI followed by a drop are seen 
at 2\farcs8 and at B0 and B1. 
We note that all these locations are also 
sites of velocity gradients (Section~\ref{spectra}) and this 
coincidence points toward a shock nature of 
the exciting mechanism, as it is dicussed in Sect. 5.2.
At B0, the HVI jet slows 
down and merges with the MVI component,  that
remains dense to beyond B1 (Fig.~\ref{bimapEC1}).

\subsection{Jet width in the initial channel}
\label{sectfwhm}

Measurements of the jet width close to its base are useful in constraining models for launching jets. 
We estimate the jet width, for the first 0\farcs7 of the flow, as the full-width-half-maximum (FWHM) of the intensity profile across the jet. The intensity profiles are obtained by using the seven slit positions to construct an image of the jet in each of four emission lines ([\ion{O}{i}]$\lambda$6563, [\ion{N}{ii}]$\lambda$6583,
[\ion{S}{ii}]$\lambda$6716, [\ion{S}{ii}]$\lambda$6731), and in three velocity intervals (similar to Fig.~\ref{bimapEC1}).  
The measured FWHM is deconvolved by 
subtracting in quadrature the FWHM of a reconstructed 
image of the stellar continuum in a wavelength interval of 3.2 \AA,
which turns out to be 13 AU (due to the overlap of the slit width).

\begin{figure}[!htb]
      \resizebox{\hsize}{!}{\includegraphics{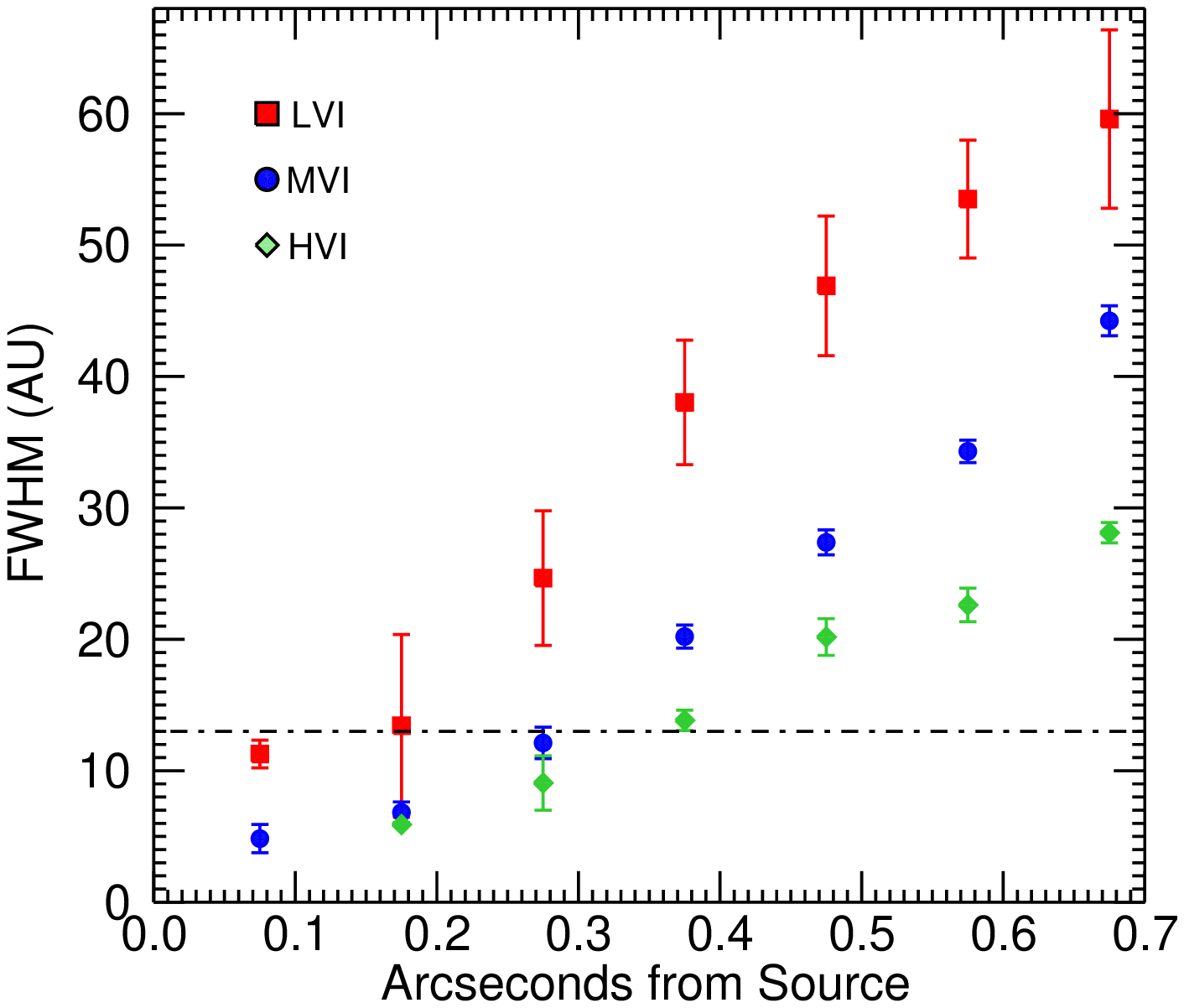}}
  \caption{DG Tau jet width 
in the first 0\farcs7. 
Each point is the average obtained from four different lines, 
with the error bars indicating the dispersion. 
Dashed line: FWHM of the stellar continuum. 
}
\vspace{-0.4cm}
  \label{fwhm}
\end{figure}

The jet width estimates are shown in Fig.~\ref{fwhm}, where each 
point is the average of the values obtained in 
the different lines in each of the velocity intervals
(with the error bar indicating the dispersion). 
The magnitude of the opening angle of the jet depends on the velocity interval, with the LVI being wider and the HVI narrower. However, within each velocity interval, the slope remains almost constant. 

Our results for the jet width are in agreement 
with the values estimated in \citet{Dougados00} and
\citet{Woitas02}, where these velocity-integrated results are close to our MVI values.
In a recent infra-red study of the DG Tau jet \cite{Agra-Amboage11}, the [\ion{Fe}{ii}] emission at velocities of -300 to -160 km s$^{-1}$ appears to have a smaller opening angle 
than our corresponding MVI emission, 
but similar to our HVI emission. This difference may be due to variability
of the flow over a six year interval. For example, 
the data in \cite{Agra-Amboage11} 
may have been taken between two episodes of intermittent inflation 
of hot plasma close to the source. 
Indeed, in contrast to our spectra, the [\ion{Fe}{ii}] emission 
did not show any emission at velocities above -300 km s$^{-1}$. 
Monitoring of [\ion{Fe}{ii}] emission over time intervals 
of 1-2 years would help clarifying this issue.

\section{Discussion} \label{discuss}

\subsection{Flow structure} 
\label{flowstr} 

The emerging picture is that of a flow initially collimated and  
characterised by smoothly varying properties, which    
soon undergoes strong perturbations represented by bright spots in the PV plots corresponding to the tips of the bow-like features described by \citet{Bacciotti00, Lavalley97}. 

The initial jet channel, up to at least position A2, has characteristics     
consistent with an 
overall onion-like kinematic structure, like the one predicted by 
classical models of jet launching, such as the so-called Disk-wind  
(e.g. \citealt{Ferreira06, Pudritz07}) or X-wind \citep{Shu00, Shang02}.
In this context an interesting aspect is 
the apparent acceleration of the 
gas between the star and A2.
Magnetocentrifugal models of disk winds predict an acceleration 
to asymptotic velocities 
on a scale proportional to the footpoint radius, $r_0$ (i.e. the radius from the star in the disk plane from which the jet is launched). 
The terminal speed is reached at about 1000$r_0$ for the solutions 
in  \citealt{Garcia01} (cfr. their Fig. 1) and at about 100$r_0$ for the solutions in \citealt{Pesenti04}. 
Our observed acceleration would then imply a launch radius of 0.3 or 3 AU, respectively. 
Alternatively, 
the ejection velocity has decreased over time, or 
the apparent acceleration is actually the effect of the  
superposition of separate components.

{Pyo03} examine the jet structure at lower spatial resolution
in [\ion{Fe}{II}] infra-red emission. Two well-separated velocity components are 
detected at epoch 10/2001. Their low radial velocity component 
of -80 km s$^{-1}$, located  at 0\farcs4, may correspond to the bright spot seen
at the stellar position in our 
LVI, while their
``high radial velocity component'' at  -220 km s$^{-1}$, located at 
0\farcs6 - 0\farcs8, may
correspond to the bright elongated feature seen in our MVI in the 
[\ion{N}{ii}] PV plots at 0 to 0\farcs4 from the star.
No HVI material is reported in \citet{Pyo03}, but 
the A1HV feature would have moved out of 
their 1\farcs6-long diagrams in two years. 

\subsection{Gas heating} 
\label{gheat} 

While our results are in agreement with previous studies, 
such as \citet{Lavalley-Fouquet00}, our analysis improves 
on these studies by quantifying the higher jet densities in the region close to the star. 

In addition, the HST resolution reveals that  
the ionisation peaks upstream A1 have different positions in 
the different velocity bins (0\farcs8 for the HVI, 
1\farcs1 for the MVI, and 1\farcs4 for LVI 
(with a smaller  peak at 0\farcs5)), 
followed by a decrease. 
This  behaviour is reminiscent of the 
excitation produced by a Disk-wind heated by
ambipolar diffusion (\citet{Garcia01}, Figs. 1, 2). In this model the slower 
outer material reaches its maximum $x_e$ at a greater distance than the inner higher velocity gas. 
The $x_e$ value, however, is much smaller than our value at the observed location. 
Alternatively, the HVI and MVI $x_e$ peaks may be due to marginally resolved 
shocks which are not evident in the forbidden emission lines. 
Indeed, in \citet{Bacciotti00}, small condensations are visible at 
0\farcs5 and 1\farcs1 but only in the strong H$\alpha$ line, 
in reconstructed images at various velocities. 
This would be in line with the conclusions of  
\citet{Lavalley-Fouquet00}, who find that the observed  
[\ion{N}{ii}]/[\ion{O}{i}] ratio are compatible only with shock heating, 
and of the numerical study of \citet{Massaglia05}, 
where the variations of $x_e$ and $n_e$ in 
distance were reproduced by a continuous series of identical 
shocks travelling along a jet of decreasing density.

We stress that we see an increase in total density $n_H$ toward each knot, 
at the same location of a sharp velocity gradient.
This is a key result as these sudden compressions indicate clearly 
that the luminous knots in the DG Tau jet are 
generated by propagating shock fronts. For example, 
at A1HV,  the high velocity 
tip of the A1 structure, there is a velocity gradient associated to 
a high $n_H$ value in the HVI.
Then at B0, the velocity jump and the local 
increase in density suggests that this knot is 
shocking the slower gas at B1. Again
feature B1 has the properties expected for a shock front 
generated by higher velocity gas catching up with slower material emitted
at an earlier time. At 4\farcs1, just downstream of the B1 emission peak, we see 
a supersonic velocity jump associated with a  gradient in density
and ionisation. 

%
The increase in total density in proximity  to velocity jumps, 
however, does not always correspond   
to an increase in ionisation.
At A1HV, for example, no peak in $x_e$ is found. 
Again at B0 one would expect an increase in ionisation, but 
$x_e$ was already high upstream of B0.
Spatial offsets between the position of peaks in $x_e$ and the position of shock fronts 
have been  found in other works conducted on similar spatial scales, see e.g. \citealt{Hartigan07} 
for the HH 30 jet.
The lack of variation in $x_e$ does not necessarily exclude 
a shock. 
In  a number of cases, in fact, line ratio modelling has shown evidence of substantial 
pre-shock ionisation (see, e.g.,  \cite{Hartigan04} in HN Tau jet and 
\cite{Tesileanu12} for the jet from RW Aur).
If a strong pre-shock ionisation was created, for example by 
the passage of a previous front, or by the 
x-ray field associated with this jet \citep{Guedel11},   
this ionisation could endure in a low density gas because of the 
slow recombination time \citep{Bacciotti99}. 
The half-life of free electrons  
is given by $ t_{\rm rec} = (n_e \alpha_H(T_e))^{-1}$, where $\alpha$ is the 
hydrogen recombination coefficient (which is weakly dependent on $T_e$).
Taking $\alpha_H = 2.5$ 10 $^{-13}$ cm$^3$ s$^{-1}$ \citep{Osterbrock89} 
and for the HVI upstream of B0 $n_e = 7~10^3$ cm$^{-3}$, 
we obtain a recombination time of $t_{\rm rec} = 6~10^8$ s. Combining this with a radial velocity 
of -300 km s$^{-1}$ and the jet inclination 
of 38$^{\circ}$ to the line of sight, we find that the free electrons can travel  
a distance of about 3$''$ in the plane of the sky before recombination. 

Finally, all PV plots show a slight asymmetry with respect to the axis.  
This jet wiggle was first detected by 
\citet{Dougados00}, and is clear in Fig.~\ref{bimapEC1} 
in the faint region between A1 and B0. 
Following \citet{Lavalley-Fouquet00}, the wiggling 
may explain the high excitation in this region, as 
a bending of the jet by only 10$^{\circ}$ at the 
observed velocities can produce oblique fronts 
with shock speeds of up to 70 km s$^{-1}$.

\subsection{Mass outflow rate} \label{mflux}

An important parameter in any jet launching model is the mass outflow rate. To ensure accuracy in estimates, it is vital to make measurements as close as possible to the base of the jet. Here it is hoped that the plasma parameters are still dominated by the physics of the launching mechanism, rather than interactions with the environment through which it propagates. Our HST/STIS dataset allows us to estimate the mass outflow rate of the jet, $\dot{M}_j$, for the first  0\farcs7  from the star. 
Inspired by the magneto-centrifugal models  (\citealt{Cabrit99}, 
\citealt{Pudritz07}, \citealt{Shang02}),
we assume that the jet flows along   
nested magnetic surfaces, with a poloidal velocity 
which decreases with distance from the jet axis.
Based on Fig.~\ref{fwhm}, 
the flow is structured in three nested cones around a hollow core, with boundary surfaces labelled 
$k=$1,2,3 and 4 according to increasing opening angle. Therefore, k=2, 3 and 4 mark 
the outer surfaces of the HVI, MVI and LVI jet cones, respectively. 
Once the jet reaches a distance, $d$, of 50 AU above the disk plane, the k=1 surface opens out to a diameter 
of either 6 AU for the disk-wind \citep{Cabrit99} or 2 AU, for the 
X-wind model \citep{Shang02}. 

At each distance, $d$, 
the jet material of each cone crosses an annulus of area 
$\pi~(r_k^2 - r_{k-1}^2) $,  
where 
$r_k(d) = FWHM(d)/2$ is the 
radius of a given cone surface.  
We define $\alpha_k$  as half of the opening angle of a given cone. From Fig.~\ref{fwhm}, 
we find that $\alpha_k =8^{\circ}, 11,^{\circ}, 19^{\circ}$ for $k=2, 3, 4$, respectively, and 
$\alpha_k < 3^{\circ}$ for k=1 if we consider the Disk-wind model. 
The mass flux $\dot{M}_{j}$ in each velocity interval, labelled by the outer cone surface $k$ 
of that interval, is then given by
\begin{equation}
\dot{M}_{j,k}  = \mu m_p f_c n_{H,k}~ |\bar{v}_{pk}| \cos{\beta_k} ~\pi~(r_k^2 - r_{k-1}^2)~~~~~~~~{\rm for}~k=2,3,4,
\label{mloss}
\end{equation}
where $n_{H,k}$ is the hydrogen density in the velocity interval region
(with values taken from Fig.\,\ref{prof1-Dnh}),
$\mu =1.41$ is the average
atomic weight per hydrogen atom \citep{Allen}, $m_p$ the proton mass,
$i = 38^{\circ}$ is the jet inclination angle 
with respect to the line of sight, and $ \beta_k=(\alpha_k+\alpha_{k-1})/2$. 
In the above expression,  
$|\bar{v}_{pk}|$ is the module of the average poloidal velocity 
in the considered interval, given as
\[
|\bar{v}_{pk}|={
\cos{i}\cos{\beta_k} ~|v_k| - 0.5\sin{i}\sin{\beta_k} ~|\delta v_k| 
\over \cos^2{i} - \sin^2{\beta_k}},
\]
$v_k$ is the  radial velocity at the middle of the interval and 
$\delta v_k$ the width of the velocity interval.  
Finally, $f_c$  is a correction factor which accounts for 
compression of the flow in unresolved shocks, estimated  
as the square root of the 
inverse of the post-shock compression \citep{Hartigan94}. 
The latter is found to be 8-10 after 10 years evolution for an 
average shock velocity around 50 km s$^{-1}$ and pre-shock 
magnetic field around 100 $\mu G$, \citep{Massaglia05}, leading to 
$f_c = 1/3$. 

\begin{figure}[!htb]
  \resizebox{\hsize}{!}{\includegraphics{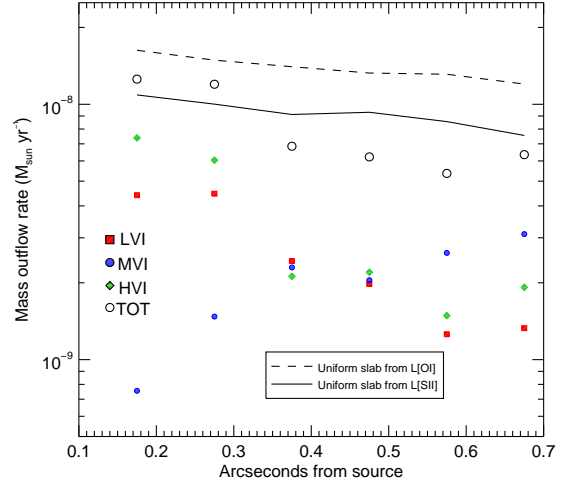}}
      \caption{Mass outflow rate of the jet, $\dot{M}_{j}$, using two methods: (i) jet density traversing annuli of nested cones, represented by data points broken down by velocity interval and also totalled; (ii) jet density obtained via emission line luminosities from a uniform slab, as in \citet{Hartigan95}, represented by curves. 
Uncertainties are about 50\% in every velocity interval. 
}
         \label{mlr}
   \end{figure}

\begin{figure}[!htb]
  \resizebox{\hsize}{!}{\includegraphics{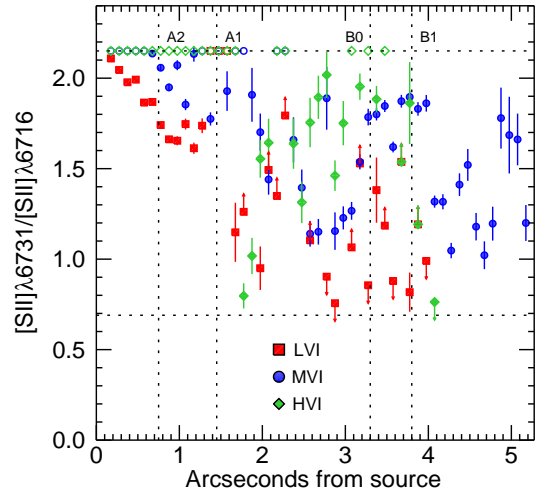}}
      \caption{
1-D profile of the [\ion{S}{ii}]31/16 ratio along the flow, derived from integration of the line 
surface brightness across the jet and over each velocity interval.    
Horizontal dashed lines indicate upper and lower density limits on validity of the ratio as a $n_e$ diagnostic.
Empty symbols mark positions and velocities for which
the modified BE-technique has been applied. 
Upper and lower limits arise when one of the lines is undetected, and so its
flux has been set to 3$\sigma$.
}
         \label{1Ds3116}
   \end{figure}

\begin{figure}[!htb]
  \resizebox{\hsize}{!}{\includegraphics{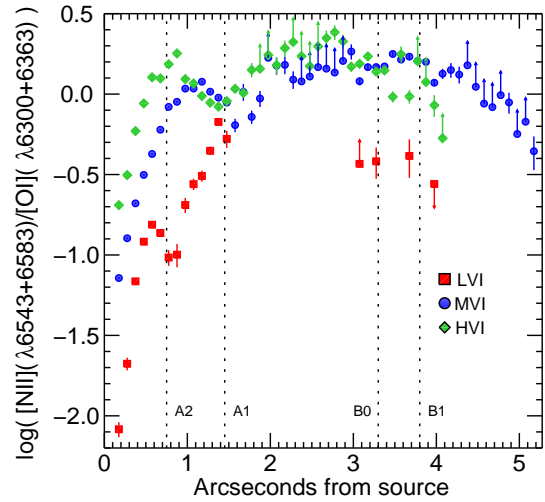}}
      \caption{
Same as Fig. \ref{1Ds3116} for the 
logarithm of 
the [\ion{N}{ii}]/[\ion{O}{i}] line ratio.
}
         \label{1Dn_o}
   \end{figure}

\begin{figure}[!htb]
  \resizebox{\hsize}{!}{\includegraphics{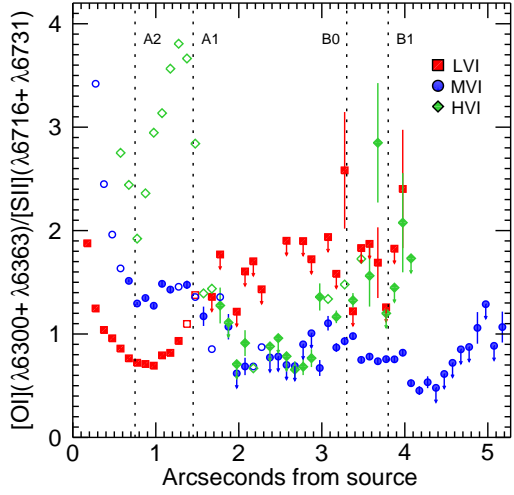}}
      \caption{
Same as Fig. \ref{1Ds3116} for the [\ion{O}{i}]/[\ion{S}{ii}] line ratio. 
Due to the prominence of [\ion{O}{i}] emission over the [\ion{S}{ii}] lines close to the star, a few points are off the scale. 
Their values are: 6.58 at 0\farcs175, for the MVI, and 12.10, 14.49, 9.54 and 4.99   
at 0\farcs175, 0\farcs275, 0\farcs375 and 0\farcs475, respectively, for 
the HVI.
}
         \label{1Do_s}
   \end{figure}

The results for $\alpha_1 = 3^{\circ}$ (i.e. the Disk-wind model) 
are presented in Fig.\,\ref{mlr}. 
Neglecting the errors on distance and inclination, the
uncertainty arises in the density and FWHM estimates.
On average the uncertainty is of  20\%, 30\%, and 40\% for the 
HVI, MVI, and LVI mass loss rates, respectively, as well as an additional 20 to 30\% error 
introduced by the modified BE-technique for the MVI and HVI densities. 
The net effect is an uncertainty of 50\% in every velocity interval.

Fig.\,\ref{mlr} shows that the contribution of the 
LVI and HVI is dominant initially,  
but it decreases with distance by 
75\%, due to the fall of density.
By contrast, the MVI contribution 
increases beyond the LVI and HVI contributions 
from  0\farcs5.
The average total mass outflow rate of the jet is
$\dot{M}_j$ = 8 $\pm$ 4 $10^{-9}$~ M$_\odot$ yr$^{-1}$,  slightly decreasing 
with distance which is probably due to flux falling 
outside the 0\farcs5 slit coverage as the jet diverges. 
Assuming a smaller central hollow cone, as in the X-wind model,
only the HVI contribution increases, to a maximum of 25\%
for $\alpha_1=0$.
Since the data are flux-calibrated, $\dot{M}_j$ could also
be estimated using the luminosity-based method of \citet{Hartigan95},
finding good agreement (Fig.\,\ref{mlr}). 

The derived $\dot{M_j}$ values of $10^{-9}$ to $10^{-8}$ match typical values
for jets from CTTSs
(e.g. for RW AUR $\dot{M}_j \sim 4.6~10^{-9}$ M$_\odot$ yr$^{-1}$ 
\citep{Melnikov09} and in CW Tau $\dot{M}_j\sim 7~10^{-9}$ M$_\odot$ yr$^{-1}$
\citep{Coffey08}), and
compare well with  previous estimates for  DG Tau. 
At 0\farcs3 from the star, \citet{Coffey08} find 
$\dot{M}_j$= 2.6 and  4.1  10$^{-8}$ M$_\odot$ yr$^{-1}$  
at -100 and -225  km s$^{-1}$. 
These values are reconciled with ours by: applying FWHM deconvolution 
(i.e. yielding jet widths of 24 and 15 AU, respectively); 
replacing the unreliable value for $x_e$ at high velocity; 
and introducing $f_c=$ 1/3. 
Meanwhile, \citet{Agra-Amboage11} obtain a value of 
$\dot{M}_j \sim$ 1.6 and 1.7 10$^{-8}$ M$_\odot$ yr$^{-1}$ at -300 and -135 km s$^{-1}$.
Their `jet density and cross-section' method
at 0\farcs4 -- 0\farcs7 (their Fig.~9)
gives results in agreement with ours, if $f_c=$ 1/3 is applied.
Finally, \citet{Lavalley-Fouquet00} reported 
$\dot{M}_j$  $\sim$ 0.2, 0.8, and 0.4 10$^{-8}$  M$_\odot$ yr$^{-1}$  
in their LVI, MVI, and HVI components 
beyond 1\farcs2 from the star, again in line with our estimates. 

Given that 
the range of accretion rates prevailing in DG Tau 
is found to be $\dot{M}_{acc} = 3 \pm 2\,10^{-7}\,M_{\odot}\,yr^{-1}$ 
\citep{Agra-Amboage11}, 
for the two-sided jet the ejection to accretion ratio  
$\dot{M}_j\,/\,\dot{M}_{acc}$ turns out to vary between 0.03$\pm$0.01 and 0.16$\pm$0.08, which is 
compatible with the range predicted by Disk-wind models 
(e.g. 0.01 $< \dot{M}_j/ \dot{M}_{acc} <$ 0.2, 
\citealt{Ferreira06}). 

\subsection{Angular momentum flux} \label{amflux}

In these spectra \citet{Bacciotti02b} 
found systematic differences in the Doppler shift 
of the LVI emission 
either side of the jet axis. 
These measurements were tentatively interpreted as the jet rotating about its axis as it propagates.  
Toroidal velocities were found to be $\sim$ 5-20 km s$^{-1}$ at a radius of 20-30 AU from
the jet axis and a distance of 40-80 AU from the star-disk plane.
Subsequently, similar differences in Doppler shift were detected 
in this and other jets \citealt{Coffey04}, 
\citealt{woitas05}, \citealt{Coffey07}). 
These results seem to support magneto-centrifugal jet launch and, in particular, the fact that jets 
extract the excess angular momentum from the star-disk system. 
Our estimate of $\dot{M}_j$ now enables us to test the extraction efficiency. 
We approximate the angular momentum of the jet in a given velocity interval 
(where k labels the outer cone surface of that interval) as follows:
\begin{equation}
\dot{L}_{j,k}\,=\,\int_{r_{k-1}}^{r_{k}}\, \mu m_p n_{H,k} v_{\phi} v_k \,2 \pi r^2\, dr
\approx  \bar{v}_{\phi, k} \dot{M}_{j,k}  \bar{r}_k
\label{momangteor}
\end{equation}
where $v_k$ and  $v_{\phi}$ are the  poloidal and toroidal components of the jet velocity. 
In DG Tau, $v_{\phi}$ could only be measured 
for LVI and MVI \citep{Bacciotti02a, Coffey07}, 
so  Equation~\ref{momangteor} simplifies to
\begin{equation}
\dot{L}_j\,=\,v_{\phi}\,(\dot{M}_{j,MVI}\,{\bar{r}_{MVI}}\,
+\,\dot{M}_{j,LVI}\,{\bar{r}_{LVI}}).
\label{momangoss}
\end{equation}
where $\overline{r}_{MVI}$ and $\overline{r}_{LVI}$ 
=($r_k$ + r$_{k-1}$)/2, for $k=3, 4$ respectively. 
Given $v_\phi \sim 15 \pm 5$ km s$^{-1}$ for DG Tau at 0\farcs3 
from the star \citep{Coffey07}, and taking the values of mass outflow rates and FWHMs 
(with their error) derived at this distance, 
we obtain 
$\dot{L}_j \sim  6.1 \pm 3.7 $ 10$^{-7}$ and $ 1.2 \pm 0.8 $ 10$^{-7}$  M$_{\odot}$ yr$^{-1}$ AU km s$^{-1}$ 
for the LVI and MVI components, respectively .
The difference with higher values given by \citet{Bacciotti02b} and \citet{Coffey08}, 
$\dot{L}_j \sim = $ 
3.8 10$^{-5}$ and 1.3 10$^{-5}$ M$_{\odot}$ yr$^{-1}$ AU km s$^{-1}$, respectively,
arises because we calculate lower mass outflow rates and smaller radii. 
\citet{Ferreira06}, however,  show that for DG Tau at 50 AU from 
the star, the outer streamlines of the wind have not yet reached the asymptotic regime 
and contain only half of their final angular momentum, contrary to the inner streamlines. 
Accounting for this effect and for the presence of a symmetric red-shifted  jet 
we obtain globally  $\dot{L}_j \sim 2.9 \pm 1.5$ 10$^{-6}$ M$_{\odot}$ yr$^{-1}$ AU km s$^{-1}$.

In order to allow accretion to proceed, the disk  must loose angular momentum. 
The excess angular momentum to be removed from the disk, 
$\dot{L}_D$, can be calculated as in \citet{woitas05}, Sect. 4.2.
We consider a Disk-wind  launched  at a broad 
range of disk radii, from the innermost region where the disk is truncated by the stellar 
magnetosphere at $r_{in} \sim$ 0.03 AU, to the outermost region at $r_{out} \sim$ 3 AU 
(the latter value is dictated by limiting poloidal velocities 
of 50 km s$^{-1}$  \citep{Bacciotti02b, Anderson03, Pesenti04}). 
The disk material falling below the inner radius is accreted in full onto the star, such that 
the disk mass flux at the inner radius  $\dot{M}_{D,in}$ is equal to  $\dot{M}_{acc}$, 
the mass accretion flux. Conservation of mass in the disk 
dictates that the mass flux in the disk at the outer radius,  $\dot{M}_{D,out}$, satisfies 
$ \dot{M}_{D,out} =  \dot{M}_{acc} + 2 \dot{M}_j $.
Considering the range of $\dot{M}_{acc}$ found in DG Tau 
and the estimate of $ \dot{M}_j $ reported in 
the previous section,  we obtain
$\dot{M}_{D,out} =$ 3.2 $\pm$ 2.1 10$^{-7}$  M$_{\odot}$ yr$^{-1}$. 
The angular momentum which the disk looses between the outer and inner radii is given by
\begin{equation} 
\dot{L}_D =  \dot{M}_{D,out}v_{K,out}r_{out} - \dot{M}_{D,in}v_{K,in}r_{in}
\end{equation}
where $v_{K}$ is the Keplerian velocity of the disk. With 
M$_{\star}$ = 0.5 M$_{\odot}$,  $v_{K}$ = 122 
and 11 km s$^{-1}$ 
at the inner and 
outer radii, respectively, and therefore 
$\dot{L}_D \sim $ 9.5 $\pm$ 7.7 ~10$^{-6}$  M$_{\odot}$ yr$^{-1}$ AU km s$^{-1}$. 
$\dot{L}_D$ and  $\dot{L}_j$ are of the same order of magnitude, and both present 
a wide range of variation. It cannot be excluded that the 
atomic jet is carrying away all the excess 
angular momentum from the disk out to 3 AU, 
if the accretion rate was 
at the low end of the range of $\dot{M}_{acc}$
measured for DG Tau over the period 1988 - 2003 \citep{Agra-Amboage11}.
Unfortunately, however, the large uncertainties still affecting the mass outflow and accretion rates  
prevent a firm conclusion on this point.

\section{Summary and conclusions} \label{concl}

We analysed a set of seven high angular resolution  
HST/STIS spectra of the first 5 arcseconds of the outflow from DG Tau, taken
in January 1999 with spectral and spatial 
resolution of $\sim 50 $ km s$^{-1}$ and 0\farcs1.
Previously published results based on this extraordinarily rich dataset include: the basic morphology of the jet 
in the first 2\arcsec $\,$ from the star 
\citep{Bacciotti00}; a preliminary set of  spectral diagnostics 
over the same region \citep{Bacciotti02a}; 
indications of jet rotation \citep{Bacciotti02b}. 
Here, we continue to exploit the 1999 HST/STIS 
dataset, to achieve a high resolution parameterisation of the jet plasma physics,
investigating it in three dimensions: along the first 5\arcsec of the jet; 
across the jet width; and in velocity space. 
We provide the PV plots of the forbidden emission lines and their ratios.  
From these we derive PV plots of 
the electron density $n_e$, hydrogen ionisation fraction $x_e$,  
and total hydrogen density $n_H$, by applying an updated version of 
the BE-technique (first published in \citet{Bacciotti99}). 
The presentation of the results as PV plots has the advantage of retaining all the 
spatio-kinematic information available. 
To assist with the interpretation, we also create  
2-D images, and 1-D profiles along the jet, of the plasma parameters
in each of three velocity intervals, 
defined as LVI from -120 to +25 km s$^{-1}$, 
MVI from -270 to -120 km s$^{-1}$ and 
HVI from -420 to -270 km s$^{-1}$,
applying the updated technique to binned data.
Our main conclusions are listed below. 

Within the first arcsecond, the flow presents  
smoothly varying kinematic properties,
with an apparent continuous acceleration in [\ion{S}{ii}] and [\ion{O}{i}] PV plots,
reminiscent of a magneto-centrifugal jet launch mechanism.
The [\ion{N}{ii}] emission is concentrated in 
medium to high velocities, and the identified features in the flow (A2, A1, B0, B1) 
are associated with supersonic velocity jumps as expected of shocked gas. 
Previous reports of an onion-like kinematic structure 
in the initial (first 1\arcsec) channel are confirmed, but 
the new estimates of 
the jet  excitation properties 
indicate that the efficiency of their ionising mechanism 
is different in the moderate-high velocity components of the flow with respect to 
the surrounding  slower, wider flow.

To find the plasma parameters  
above the high density limit of [\ion{S}{ii}], the modified BE-technique was applied. 
At the beginning of the jet, values up to $n_e \sim$ 10$^5$ cm$^{-3}$ were found 
for MVI and HVI, while lower values of $n_e \sim$ 10$^4$ cm$^{-3}$ were found in the LVI. 
In the same region, $x_e$ increases markedly in the MVI and HVI, 
up to values of 0.7 and 0.6, respectively, close to A2.
By contrast, the LVI value remains low i.e. $x_e \leq$ 0.3 within the location of A1. 
This  suggests a fundamental difference between 
the dominant ionisation process 
in the MVI and HVI components 
with respect to the LVI component. 
Proceeding along the jet beyond feature A1, 
$n_e$ decreases in all velocity intervals. 
Meanwhile, $x_e$ is found to remain high (0.7-0.8) 
in the MVI and HVI (the LVI is not visible here), both in the 
A2-B0 stream, and between features B0 and B1. 
After B1, $n_e$ and $x_e$ can only 
be determined for the MVI, and both decrease 
to low values, along the jet to 5 \arcsec. 
The total hydrogen density $n_H$ for the three velocity intervals 
is similar within a factor 3-4 all along the jet. Beyond A1
we have no $n_H$ estimates for the LVI, but values similar to our HVI
are found here by  \cite{Lavalley-Fouquet00}. 
Overall, the total density
drops in magnitude over 4 orders along the jet (from almost 10$^7$ to 10$^3$ cm$^{-3}$). 
Our results are in agreement 
with previous determinations of the excitation parameters, in the regions where the comparison was feasible. 

Remarkably, our analysis shows absence of significant variations in the plasma parameters 
across the jet width in each velocity interval. 
Slightly downstream the emission peaks at AHV1, B0 and B1, and in the HVI, 
local pronounced gradients of the total density  are found  to be 
coincident with velocity jumps. 
This is direct evidence that the  gas is compressed locally at the position 
of sharp velocity gradients, supporting a shock origin for the observed knots,
as proposed by \citet{Lavalley-Fouquet00}. This conclusion holds true 
even if not in all cases an 
increase in ionisation is detected at the velocity jumps, as it is 
in the MVI at B1.
In fact the absence of variation in the ionisation level 
does not exclude a shock, as the ionisation created by the 
passage of a previous shock or by a radiation field could endure because of the 
slow recombination time in a rarefied medium \citep{Bacciotti99}.  
We confirm reports of wiggling  between A1 and B0, 
supporting the suggestion by \citet{Lavalley-Fouquet00}, that the 
high excitation values in this region may be maintained via the occurence of lateral shocks. 

The mass outflow rate of the jet, $\dot{M_j}$, is found as a function of  
velocity and distance from the star, in the initial portion of the jet (0\farcs1 - 0\farcs7), 
using the jet density and cross-section measurements. 
$\dot{M_j}$ is similar in the HVI and LVI, and decreases along the flow. 
Meanwhile,  the MVI value is initially lower, but then increases 
with distance until it  dominates after 0\farcs5. 
The total mass flux, of all velocity intervals, is on average
$\dot{M_j} \sim$ 8$\pm$4~10$^{-9}$ M$_\odot$ yr$^{-1}$, and it is found to 
decrease slightly with distance, 
but this could be a bias introduced by  the line flux loss at the slit borders. 
Results were cross-checked with mass outflow rates obtained via 
emission line luminosity (cfr. \citet{Hartigan95}), finding good agreement.
Taking into account differences in the derivation procedure, 
our results are also in agreement with previous
estimates  for this jet, and in the typical range for CTTS jets. 
Given the range of mass accretion rates found  in DG Tau of 
$\dot{M}_{acc}\,\sim \, (3 \pm 2) \,10^{-7}$\,M$_{\odot}$\,yr$^{-1}$ 
\citep{Agra-Amboage11}, the ratio of mass ejection to mass accretion,
$\dot{M}_j\,/\,\dot{M}_{acc}$, for the supposedly simmetric bipolar jet 
can vary between  0.03$\pm$0.01 and 0.16$\pm$0.08, which is 
compatible with the range predicted by Disk-wind models.

Combining the derived mass outflow rates with previously published  
toroidal velocities for the LVI and MVI material at 0\farcs3 form the star, 
we estimate the angular momentum transported by these components.
Considering  two symmetric jet lobes and allowing 
a correction for 
the fraction of the angular momentum  still in the disk-wind magnetic field 
before the asymptotic regime is reached,
$\dot{L}_j$ turns out to be $ \sim$ (2.9 $\pm$ 1.5) 10$^{-6}$ M$_{\odot}$ yr$^{-1}$ AU km s$^{-1}$.
We tentatively compare this estimate to the amount of angular momentum lost by the disk to allow 
accretion, $\dot{L}_D$. Proceeding as in \citet{woitas05}, 
we find $\dot{L}_D \sim $ 9.5 $\pm$ 7.7 ~10$^{-6}$  M$_{\odot}$ yr$^{-1}$ AU km s$^{-1}$, 
which indicates that the two quantities are of the same order of magnitude, 
and comparable if the accretion rate was at its lower values when the 
material probed by our data was ejected. The large uncertainties affecting both estimates, 
however, prevent further conclusions. 

In summary, the physical structure of the DG Tau jet 
reveals patterns of variation in parameters which are expected of magneto-centrifugal jet launch models. 
However, the situation is complicated by the simultaneous presence 
of other features, like hints of a multi-component flow, and shock fronts formed on different temporal scales,
which seem to reach beyond this simple scenario. 
The presented plasma maps constitute a powerful benchmark for testing new alternatives. 

\begin{acknowledgements} 
The authors wish to thank the referee, Sylve Cabrit, for her detailed and 
thorough reports, that led to a significant improvement in 
the derivation and presentation of the results. 
L.M. thanks S. Cabrit and C. Dougados for the hospitality at IAP during the 
preliminary analysis of the data.
T.P.R. acknowledges support from Science Foundation Ireland under 
grant 07/RFP/PHYF790. 
\end{acknowledgements}

\bibliographystyle{aa}
\bibliography{biblio_1}

\begin{thebibliography}{52}
\expandafter\ifx\csname natexlab\endcsname\relax\def\natexlab#1{#1}\fi

\bibitem[{{Agra-Amboage} {et~al.}(2011){Agra-Amboage}, {Dougados}, {Cabrit}, \&
  {Reunanen}}]{Agra-Amboage11}
{Agra-Amboage}, V., {Dougados}, C., {Cabrit}, S., \& {Reunanen}, J. 2011, \aap,
  532, A59

\bibitem[{Allen \& Cox(2001)}]{Allen}
Allen, C.~W. \& Cox, A.~N. 2001, Allen's astrophysical quantities, 4th edn. (A.
  N. Cox (Springer)), 729

\bibitem[{{Anderson} {et~al.}(2003){Anderson}, {Li}, {Krasnopolsky}, \&
  {Blandford}}]{Anderson03}
{Anderson}, J.~M., {Li}, Z.-Y., {Krasnopolsky}, R., \& {Blandford}, R.~D. 2003,
  \apjl, 590, L107

\bibitem[{{Asplund} {et~al.}(2005){Asplund}, {Grevesse}, \&
  {Sauval}}]{Asplund05}
{Asplund}, M., {Grevesse}, N., \& {Sauval}, A.~J. 2005, in Astronomical Society
  of the Pacific Conference Series, Vol. 336, Cosmic Abundances as Records of
  Stellar Evolution and Nucleosynthesis, ed. T.~G. {Barnes}, III \& F.~N.
  {Bash}, 25


\bibitem[{{Bacciotti} \& {Eisl{\"o}ffel}(1999){Bacciotti} \& {Eisl{\"o}ffel}}]{BE99}
{Bacciotti}, F. \& {Eisl{\"o}ffel}, J. 1999, \aap, 342, 717

\bibitem[{{Bacciotti} {et~al.}(1999){Bacciotti}, {Eisl{\"o}ffel}, \&
  {Ray}}]{Bacciotti99}
{Bacciotti}, F., {Eisl{\"o}ffel}, J., \& {Ray}, T.~P. 1999, \aap, 350, 917

\bibitem[{{Bacciotti} {et~al.}(2000){Bacciotti}, {Mundt}, {Ray},
  {Eisl{\"o}ffel}, {Solf}, \& {Camezind}}]{Bacciotti00}
{Bacciotti}, F., {Mundt}, R., {Ray}, T.~P., {et~al.} 2000, \apjl, 537, L49

\bibitem[{{Bacciotti}(2002)}]{Bacciotti02a}
{Bacciotti}, F. 2002, in Revista Mexicana de Astronomia y Astrofisica
  Conference Series, Vol.~13, Revista Mexicana de Astronomia y Astrofisica
  Conference Series, ed. W.~J. {Henney}, W.~{Steffen}, L.~{Binette}, \&
  A.~{Raga}, 8--15


\bibitem[{{Bacciotti} {et~al.}(2002){Bacciotti}, {Ray}, {Mundt},
  {Eisl{\"o}ffel}, \& {Solf}}]{Bacciotti02b}
{Bacciotti}, F., {Ray}, T.~P., {Mundt}, R., {Eisl{\"o}ffel}, J., \& {Solf}, J.
  2002, \apj, 576, 222

\bibitem[{{Bally} {et~al.}(2007){Bally}, {Reipurth}, \& {Davis}}]{Bally07}
{Bally}, J., {Reipurth}, B., \& {Davis}, C.~J. 2007, Protostars and Planets V,
  215

\bibitem[{{Berrington} \& {Burke}(1981)}]{Berrington81}
{Berrington}, K.~A. \& {Burke}, P.~G. 1981, \planss, 29, 377

\bibitem[{{Cabrit} {et~al.}(1999){Cabrit}, {Ferreira}, \& {Raga}}]{Cabrit99}
{Cabrit}, S., {Ferreira}, J., \& {Raga}, A.~C. 1999, \aap, 343, L61

\bibitem[{{Cabrit} {et~al.}(1999)}]{Cabrit09}{Cabrit}, S.,  in 
Protostellar Jets in Context, by K. Tsinganos, T. Ray, M. Stute. 
\apss\,Proceedings series. Berlin: Springer, 2009, pp.247-257

\bibitem[{{Cerqueira} {et~al.}(2006){Cerqueira}, {Vel{\'a}zquez}, {Raga},
  {Vasconcelos}, \& {de Colle}}]{Cerqueira2006}
{Cerqueira}, A.~H., {Vel{\'a}zquez}, P.~F., {Raga}, A.~C., {Vasconcelos},
  M.~J., \& {de Colle}, F. 2006, \aap, 448, 231

\bibitem[{{Coffey} {et~al.}(2008){Coffey}, {Bacciotti}, \& {Podio}}]{Coffey08}
{Coffey}, D., {Bacciotti}, F., \& {Podio}, L. 2008, \apj, 689, 1112

\bibitem[{{Coffey} {et~al.}(2007){Coffey}, {Bacciotti}, {Ray}, {Eisl{\"o}ffel},
  \& {Woitas}}]{Coffey07}
{Coffey}, D., {Bacciotti}, F., {Ray}, T.~P., {Eisl{\"o}ffel}, J., \& {Woitas},
  J. 2007, \apj, 663, 350

\bibitem[{{Coffey} {et~al.}(2004){Coffey}, {Bacciotti}, {Woitas}, {Ray}, \&
  {Eisl{\"o}ffel}}]{Coffey04}
{Coffey}, D., {Bacciotti}, F., {Woitas}, J., {Ray}, T.~P., \& {Eisl{\"o}ffel},
  J. 2004, \apss, 292, 553

\bibitem[{{DeColle} {et~al.}(2010){De Colle}, F. {del Burgo}, C. \& {Raga}, A.C.}]{DeColle10}
{De Colle}, F. {del Burgo}, C. \& {Raga}, A.C. 2010, \apj,
  721, 929

\bibitem[{{Dougados} {et~al.}(2000){Dougados}, {Cabrit}, {Lavalley}, \&
  {M{\'e}nard}}]{Dougados00}
{Dougados}, C., {Cabrit}, S., {Lavalley}, C., \& {M{\'e}nard}, F. 2000, \aap,
  357, L61

\bibitem[{{Edwards}(2009)}]{Edwards09}
{Edwards}, S. 2009, in American Institute of Physics Conference Series, Vol.
  1094, 15th Cambridge Workshop on Cool Stars, Stellar Systems, and the Sun,
  ed. E.~{Stempels}, 29--38

\bibitem[{{Eisl{\"o}ffel} \& {Mundt}(1998)}]{Eisloffel98}
{Eisl{\"o}ffel}, J. \& {Mundt}, R. 1998, \aj, 115, 1554

\bibitem[{{Ferreira} {et~al.}(2006){Ferreira}, {Dougados}, \&
  {Cabrit}}]{Ferreira06}
{Ferreira}, J., {Dougados}, C., \& {Cabrit}, S. 2006, \aap, 453, 785

\bibitem[{{Garcia} {et~al.}(2001){Garcia}, {Cabrit}, {Ferreira}, \&
  {Binette}}]{Garcia01}
{Garcia}, P.~J.~V., {Cabrit}, S., {Ferreira}, J., \& {Binette}, L. 2001, \aap,
  377, 609

\bibitem[{{G{\"u}del} {et~al.}(2011){G{\"u}del}, {Audard}, {Bacciotti}, {Bary},
  {Briggs}, {Cabrit}, {Carmona}, {Codella}, {Dougados}, {Eisl{\"o}ffel},
  {Gueth}, {G{\"u}nther}, {Herczeg}, {Kundurthy}, {Matt}, {Mutel}, {Ray},
  {Schmitt}, {Schneider}, {Skinner}, \& {van Boekel}}]{Guedel11}
{G{\"u}del}, M., {Audard}, M., {Bacciotti}, F., {et~al.} 2011, in Astronomical
  Society of the Pacific Conference Series, Vol. 448, Astronomical Society of
  the Pacific Conference Series, ed. C.~{Johns-Krull}, M.~K. {Browning}, \&
  A.~A. {West}, 617

\bibitem[{{G{\"u}del} {et~al.}(2008){G{\"u}del}, {Skinner}, {Audard}, {Briggs},
  \& {Cabrit}}]{Guedel08}
{G{\"u}del}, M., {Skinner}, S.~L., {Audard}, M., {Briggs}, K.~R., \& {Cabrit},
  S. 2008, \aap, 478, 797


\bibitem[{{Hartigan} {et~al.}(1994){Hartigan}, {Morse}, \&
  {Raymond}}]{Hartigan94}
{Hartigan}, P., {Morse}, J.~A., \& {Raymond}, J. 1994, \apj, 436, 125


\bibitem[{{Hartigan} {et~al.}(1995){Hartigan}, {Edwards}, \&
  {Ghandour}}]{Hartigan95}
{Hartigan}, P., {Edwards}, S., \& {Ghandour}, L. 1995, \apj, 452, 736

\bibitem[{{Hartigan} {et~al.}(2004){Hartigan}, {Edwards}, \&
  {Pierson}}]{Hartigan04}
{Hartigan}, P., {Edwards}, S., \& {Pierson}, R. 2004, \apj, 609, 261

\bibitem[{{Hartigan} \& {Morse}(2007)}]{Hartigan07}
{Hartigan}, P. \& {Morse}, J. 2007, \apj, 660, 426


\bibitem[{{Hartmann} \& {Raymond}(1989)}]{Hartmann89}
{Hartmann}, L. \& {Raymond}, J.~C. 1989, \apj, 337, 903

\bibitem[{{Hudson} \& {Bell}(2005)}]{Hudson05}
{Hudson}, C.~E. \& {Bell}, K.~L. 2005, \aap, 430, 725

\bibitem[{{Keenan} {et~al.}(1996){Keenan}, {Aller}, {Bell}, {Hyung}, {McKenna},
  \& {Ramsbottom}}]{Keenan96}
{Keenan}, F.~P., {Aller}, L.~H., {Bell}, K.~L., {et~al.} 1996, \mnras, 281,
  1073

\bibitem[{{Kwan} \& {Tademaru}(1988)}]{Kwan88}
{Kwan}, J. \& {Tademaru}, E. 1988, \apjl, 332, L41

\bibitem[{{Lavalley} {et~al.}(1997){Lavalley}, {Cabrit}, {Dougados}, {Ferruit},
  \& {Bacon}}]{Lavalley97}
{Lavalley}, C., {Cabrit}, S., {Dougados}, C., {Ferruit}, P., \& {Bacon}, R.
  1997, \aap, 327, 671

\bibitem[{{Lavalley-Fouquet} {et~al.}(2000){Lavalley-Fouquet}, {Cabrit}, \&
  {Dougados}}]{Lavalley-Fouquet00}
{Lavalley-Fouquet}, C., {Cabrit}, S., \& {Dougados}, C. 2000, \aap, 356, L41

\bibitem[{{Massaglia} {et~al.}(2005){Massaglia}, {Mignone}, \&
  {Bodo}}]{Massaglia05}
{Massaglia}, S., {Mignone}, A., \& {Bodo}, G. 2005, \aap, 442, 549

\bibitem[{{McGroarty} {et~al.}(2007){McGroarty}, {Ray}, \&
  {Froebrich}}]{McGroarty07}
{McGroarty}, F., {Ray}, T.~P., \& {Froebrich}, D. 2007, \aap, 467, 1197

\bibitem[{{Melnikov} {et~al.}(2008){Melnikov}, {Woitas}, {Eisl{\"o}ffel},
  {Bacciotti}, {Locatelli}, \& {Ray}}]{Melnikov08}
{Melnikov}, S., {Woitas}, J., {Eisl{\"o}ffel}, J., {et~al.} 2008, \aap, 483,
  199

\bibitem[{{Melnikov} {et~al.}(2009){Melnikov}, {Eisl{\"o}ffel}, {Bacciotti},
  {Woitas}, \& {Ray}}]{Melnikov09}
{Melnikov}, S.~Y., {Eisl{\"o}ffel}, J., {Bacciotti}, F., {Woitas}, J., \&
  {Ray}, T.~P. 2009, \aap, 506, 763

\bibitem[{{Mendoza}(1983)}]{Mendoza83}
{Mendoza}, C. 1983, in IAU Symposium, Vol. 103, Planetary Nebulae, ed. D.~R.
  {Flower}, 143--172

\bibitem[{{Mundt} \& {Fried}(1983)}]{Mundt83}
{Mundt}, R. \& {Fried}, J.~W. 1983, \apjl, 274, L83

\bibitem[{{Osterbrock}(1989)}]{Osterbrock89}
{Osterbrock},  D.E., 1989, Astrophysics of Gaseous Nebulae and Active
Galactic Nuclei. University Science Books, Mill Valley, CA


\bibitem[{{Pesenti} {et~al.}(2004){Pesenti}, {Dougados}, {Cabrit}, {Ferreira},
  {Casse}, {Garcia}, \& {O'Brien}}]{Pesenti04}
{Pesenti}, N., {Dougados}, C., {Cabrit}, S., {et~al.} 2004, \aap, 416, L9

\bibitem[{{Podio} {et~al.}(2006){Podio}, {Bacciotti}, {Nisini},
  {Eisl{\"o}ffel}, {Massi}, {Giannini}, \& {Ray}}]{Podio06}
{Podio}, L., {Bacciotti}, F., {Nisini}, B., {et~al.} 2006, \aap, 456, 189

\bibitem[{{Podio} {et~al.}(2011){Podio}, {Eisl{\"o}ffel}, {Melnikov}, {Hodapp},
  \& {Bacciotti}}]{Podio11}
{Podio}, L., {Eisl{\"o}ffel}, J., {Melnikov}, S., {Hodapp}, K.~W., \&
  {Bacciotti}, F. 2011, \aap, 527, A13

\bibitem[{{Pudritz} {et~al.}(2007){Pudritz}, {Ouyed}, {Fendt}, \&
  {Brandenburg}}]{Pudritz07}
{Pudritz}, R.~E., {Ouyed}, R., {Fendt}, C., \& {Brandenburg}, A. 2007,
  Protostars and Planets V, 277

\bibitem[{{Pyo} {et~al.}(2003){Pyo}, {Kobayashi}, {Hayashi}, {Terada}, {Goto},
  {Takami}, {Takato}, {Gaessler}, {Usuda}, {Yamashita}, {Tokunaga}, {Hayano},
  {Kamata}, {Iye}, \& {Minowa}}]{Pyo03}
{Pyo}, T.-S., {Kobayashi}, N., {Hayashi}, M., {et~al.} 2003, \apj, 590, 340

\bibitem[{{Ray} {et~al.}(2007){Ray}, {Dougados}, {Bacciotti}, {Eisl{\"o}ffel},
  \& {Chrysostomou}}]{Ray07}
{Ray}, T., {Dougados}, C., {Bacciotti}, F., {Eisl{\"o}ffel}, J., \&
  {Chrysostomou}, A. 2007, Protostars and Planets V, 231

\bibitem[{{Raymond} {et~al.}(1994){Raymond}, {Morse}, {Hartigan}, {Curiel}, \&
  {Heathcote}}]{Raymond94}
{Raymond}, J.~C., {Morse}, J.~A., {Hartigan}, P., {Curiel}, S., \& {Heathcote},
  S. 1994, \apj, 434, 232

\bibitem[{{Safier}(1993)}]{Safier93}
{Safier}, P.~N. 1993, \apj, 408, 148

\bibitem[{{Shang} {et~al.}(2002){Shang}, {Glassgold}, {Shu}, \&
  {Lizano}}]{Shang02}
{Shang}, H., {Glassgold}, A.~E., {Shu}, F.~H., \& {Lizano}, S. 2002, \apj, 564,
  853

\bibitem[{{Shu} {et~al.}(2000){Shu}, {Najita}, {Shang}, \& {Li}}]{Shu00}
{Shu}, F.~H., {Najita}, J.~R., {Shang}, H., \& {Li}, Z.-Y. 2000, Protostars and
  Planets IV, 789

\bibitem[{{Takami} {et~al.}(2002){Takami}, {Chrysostomou}, {Bailey},
  {Gledhill}, {Tamura}, \& {Terada}}]{Takami02}
{Takami}, M., {Chrysostomou}, A., {Bailey}, J., {et~al.} 2002, \apjl, 568, L53

\bibitem[{{Te{\c s}ileanu} {et~al.}(2012){Te{\c s}ileanu}, {Mignone},
  {Massaglia}, \& {Bacciotti}}]{Tesileanu12}
{Te{\c s}ileanu}, O., {Mignone}, A., {Massaglia}, S., \& {Bacciotti}, F. 2012,
  \apj, 746, 96

\bibitem[{{Whelan} {et~al.}(2005){Whelan}, {Ray}, {Bacciotti}, {Natta},
  {Testi}, \& {Randich}}]{Whelan05}
{Whelan}, E.~T., {Ray}, T.~P., {Bacciotti}, F., {et~al.} 2005, \nat, 435, 652

\bibitem[{{Woitas} {et~al.}(2005){Woitas}, {Bacciotti}, {Ray}, {Marconi},
  {Coffey}, \& {Eisl{\"o}ffel}}]{woitas05}
{Woitas}, J., {Bacciotti}, F., {Ray}, T.~P., {et~al.} 2005, \aap, 432, 149

\bibitem[{{Woitas} {et~al.}(2002){Woitas}, {Ray}, {Bacciotti}, {Davis}, \&
  {Eisl{\"o}ffel}}]{Woitas02}
{Woitas}, J., {Ray}, T.~P., {Bacciotti}, F., {Davis}, C.~J., \&
  {Eisl{\"o}ffel}, J. 2002, \apj, 580, 336

\end{thebibliography}

\newpage

{\bf ON-LINE MATERIAL}


\end{document}